\documentclass[10pt,journal,compsoc]{IEEEtran}
\usepackage{cite}
\usepackage{amsmath,amssymb,amsfonts}
\usepackage{algorithm}
\usepackage{algorithmicx}
\usepackage{graphicx}
\usepackage{textcomp}
\usepackage{xcolor}
\usepackage{subcaption}
\usepackage{booktabs}
\usepackage{multirow}
\usepackage{algpseudocode}
\usepackage{ulem}
\usepackage{diagbox}
\usepackage[justification=centering]{caption}

\begin{document}

\title{CARE: Ensemble Adversarial Robustness Evaluation Against Adaptive Attackers for Security Applications}

\author{Hangsheng Zhang,
        Jiqiang Liu,
        Jinsong Dong}

\IEEEtitleabstractindextext{%
\begin{abstract}

Ensemble defenses, are widely employed in various security-related applications to enhance model  performance and robustness.
The widespread adoption of these techniques also raises many questions:
Are general ensembles defenses guaranteed to be more robust than individuals?
Will stronger adaptive attacks defeat existing ensemble defense strategies as the cybersecurity arms race progresses?
Can ensemble defenses achieve adversarial robustness to different types of attacks simultaneously and resist the continually adjusted adaptive attacks?
Unfortunately, these critical questions remain unresolved as there are no platforms for comprehensive evaluation of ensemble adversarial attacks and defenses in the cybersecurity domain.

In this paper, we propose a general \emph{C}ybersecurity \emph{A}dversarial \emph{R}obustness \emph{E}valuation (CARE) platform aiming to bridge this gap.
First of all, CARE comprehensively evaluates the defense effectiveness of model ensembles and ensemble defenses across the machine learning space for security applications.
And then, we explore adaptive ensemble attacks in two cases, including an ensemble of multiple attack methods and a transfer ensemble attack to obtain an adversarial example that satisfies the problem-space constraints.
Moreover, we propose an improved ensemble defense method: robust ensemble adversarial training, which enables the adversarial training to resist multiple attack methods simultaneously, ensuring resistance to adaptive ensemble attacks.
Finally, we conduct experiments on five security application datasets against 12 security detectors using 15 attack and 8 defense strategies.
Experimental results show that existing model ensembles and ensemble adversarial training methods fail to guarantee enhanced model robustness against adaptive attackers, whereas our robust ensemble adversarial training is capable of resisting multiple attack methods and adaptive attacks simultaneously.

\end{abstract}

}
\maketitle

\section{introduction}
Many detection systems based on artificial intelligence (AI) have been proposed in security tasks such as network intrusion detections (NIDSs) \cite{mirsky2018kitsune,marteau2021random,chen2021adsim}, malware detections \cite{vasan2020image,rabadi2020advanced} and spam detections \cite{kwon2017domain}.
Nevertheless, numerous studies have revealed that AI-based systems are vulnerable to adversarial attacks \cite{han2020practical,nasr2021defeating,ilyas2019adversarial}.
Thus, in a highly security-sensitive setting such as cybersecurity, it is of utmost importance to ensure that the AI-based security detectors remain reliable and robust in the face of adversarial attacks \cite{biggio2018wild,rosenberg2021adversarial,apruzzese2021modeling,ling2021adversarial}.

There is less research on the adversarial robustness evaluation of ensemble models than linear and neural networks models \cite{ling2019deepsec,papernot2016cleverhans,rauber2017foolbox}, despite the fact that many security-related applications use ensemble learning techniques because of their flexibility, resilience and competitive performance \cite{rong2020malfinder,zhang2020dynamic,rabadi2020advanced}.
Moreover, some studies showed model ensembles \cite{biggio2010multiple,smutz2016tree} and ensemble defenses \cite{tramer2017ensemble,tramer2019adversarial} will also enhance the robustness of models, while other researchers have challenged this claim\cite{zhang2018gradient,zhang2020decision,kariyappa2019improving}.

The controversy leads to the first question: \textit{Are general model ensembles and ensembles defenses guaranteed to be more robust than individuals?}
Existing ensemble evaluation methods \cite{shu2022omni,li2020adversarial} are insufficient to answer this question primarily due to three aspects:
(i) they are largely gradient-based \cite{rosenberg2021adversarial,shu2022omni,li2020adversarial} without systematically studying gradient-free methods, which are more suitable for cybersecurity discrete data.
(ii) they focus on homogeneous ensembles that use deep \cite{shu2022omni,li2020adversarial} and tree models \cite{chen2021cost}, but neglecte heterogeneous ensembles that combine deep and tree models.
(iii) they explore the \textit{ensemble of multiple attack methods} \cite{li2020adversarial,tramer2019adversarial,maini2020adversarial} more, and rarely examine the ensemble of perturbations from other models \textit{(transfer ensemble attack)} \cite{tramer2019adversarial}.

Second, security detectors face an arms race \cite{li2021arms} in which attackers are constantly adapting to the ensemble defense system.
As a result, a more practical attack on security detectors involves an \textit{adaptive ensemble attack (Adp-EA)} \cite{he2017adversarial,dou2020robust,tramer2020adaptive} that understand the defenses of the model and can adjust their attacks accordingly.
As a stronger adversary than static adversary, we are required to know:
\textit{Will stronger adaptive attacks defeat existing ensemble defense strategies as the cybersecurity arms race progresses?}
However, the current robustness evaluations of ensemble defense systems generally focus on single or simple ensemble attacks \cite{li2020adversarial,tramer2017ensemble,tramer2019adversarial,maini2020adversarial} without assessing the adaptability of attackers.

Third, existing defense methods such as naive adversarial training (NAT) \cite{goodfellow2014explaining,madry2017towards} are typically tailored to one type of attack, which leaves them vulnerable to other types of adversarial attacks, meaning that it has poor \textit{robust generalization} \cite{tramer2019adversarial} ability.
Thus, \textit{Can ensemble defenses (here is ensemble adversarial training) achieve adversarial robustness to different types of attacks simultaneously and resist the continually adjusted adaptive attacks?}
Although improved ensemble adversarial training methods \cite{tramer2019adversarial,tramer2017ensemble} such as average strategy or worst-case strategy as a defensive strategy have a certain level of robust generalization ability, it is unable to resist the attacker's constantly adapting strategy.

To accurately answer the above questions, we urgently need to develop an ensemble adversarial robustness evaluation platform in the cybersecurity domain.
Therefore, we propose a first general \emph{C}ybersecurity \emph{A}dversarial \emph{R}obustness \emph{E}valuation (CARE) platform for comprehensively evaluating the robustness of various model ensembles and ensemble defenses.
To be practically useful, our platform satisfies a set of desiderata as follows:
\begin{itemize}
       \item Uniform – It provides a unified interface for classic machine learning and deep learning to evaluate \textit{heterogeneous ensemble models} and \textit{transfer ensemble attacks};
       \item Comprehensive – It supports comprehensive assessment of ML/DL vulnerabilities in security applications using a variety of gradient-based and gradient-free methods;
       \item Practical – It enables more realistic \textit{adaptive attacks}, and a number of problem-space constraints in the security domain to ensure that realistic adversarial examples (AEs) can be produced;
       \item Extensible – It is easily extensible to incorporate state-of-the-art attack/defense methods and new problem-space constraints.
\end{itemize}

Specifically, CARE incorporates 15 state-of-the-art adversarial attacks, 8 defense methods, and 12 ML-based security detectors.
Adversarial attacks can be divided into three categories: 6 gradient-based attacks, 6 gradient-free attacks, and 3 ensemble attacks.
We implement 12 security classifiers, both traditional machine learning and deep learning models (such as \textit{malimg} and \textit{transformer}), along with three types of ensemble models.
These three ensemble methods and five adversarial training (AT) methods make up our adversarial defense library.
Finally, we implement a series of remapping functions to project the adversarial examples (AEs) into the real space, and we also perform an end-to-end attack against malware detectors to demonstrate our framework can scale up to a real-world problem-space attack.

The main contributions of this paper are summarized as follows:
\begin{enumerate}
    \item  We design a first general Cybersecurity Adversarial Robustness Evaluation (CARE) framework to evaluate the effect of model ensembles and ensemble defenses on the entire machine learning space.
           This framework achieves high scalability, implements a unified interface for machine learning and deep learning, supports the state-of-the-art attack and defense algorithms, more importantly, represents problem-space constraints in the security domain by remapping functions.
    \item  We present \textit{adaptive ensemble attacks} in two cases: an \textit{ensemble of multiple attack} methods and a \textit{transfer ensemble attack} to meet the measurement requirements for adaptive attacks in security scenarios.
    \item  We propose a novel adversarial defense method called \textit{robust ensemble adversarial training}, which uses Bayesian optimization to find the most robust ensemble weights and can withstand multiple attack methods and adaptive attacks simultaneously.
    \item We extensively evaluate five security datasets including a NIDSs dataset (CICIDS2017) \cite{sharafaldin2018toward}, two malware datasets (Bodmas\cite{yang2021bodmas} and PEData), a encrypted malware traffic dataset (CICAndMal2017)\cite{lashkari2018toward}, and a spam dataset (TwitterSpam)\cite{kwon2017domain}. 
              Through this systematic study, we obtain a set of interesting and insightful findings that can answer the above questions:
              (i) general model ensembles and ensemble defenses cannot guarantee the robustness of the model.
              (ii) An \textit{adaptive attack} can easily defeat a simple ensemble defense;
              (iii) our \textit{robust ensemble adversarial training} is able to resist multiple attack methods simultaneously as well as \textit{adaptive attacks}.
  \end{enumerate}    
\section{background and related work}
In this section, we first describe some other adversarial robustness evaluation platforms.
We then introduce the detectors for security applications based on ensemble learning.
Following that, we discuss some adversarial attacks against security detectors.
Finally, we review some work related to using ensemble techniques in defense scenarios, including model ensembles and ensemble adversarial training.

\subsection{Adversarial Robustness Evaluation Platform}
A few platforms are evaluating deep learning and image domain, like Cleverhans \cite{papernot2016cleverhans}, Deepsec \cite{ling2019deepsec}, Foolbox \cite{rauber2017foolbox} and ART \cite{ART}, but they do not take security domain knowledge into account, and cannot be directly transferred to cybersecurity.
As a platform in the cybersecurity domain, secml-malware \cite{demetrio2021secml} can only evaluate the robustness of Windows malware detection, and there are few detection and attack algorithms, so it is not suitable to evaluate the adversarial robustness of ensemble technologies.
In general, these platforms are not designed to ensemble and adaptive robust evaluation techniques in the cybersecurity domain.

\subsection{Ensemble Learning-based Security Detectors}
Security applications often rely on ensemble learning because of its flexibility, resilience, and competitive performance.
There are three ensemble-based approaches for security application: (i) homogeneous tree ensemble (Boosting and Bagging) (ii) deep ensemble, and (iii) heterogeneous ensemble.

For type (i), Rong et al. \cite{rong2020malfinder} and Rabadi et al.\cite{rabadi2020advanced} evaluated the performance of different ensemble tree models (boosting and bagging) for malicious traffic detection and malware detection, respectively.
By improving the basic ensemble bagging model, \cite{marteau2021random} presented DiFF-RF, an ensemble approach based on random partitioning binary trees that can detect point-wise and collective anomalies.

For type (ii), as a classic example, Mirsky et al.\cite{mirsky2018kitsune} developed KitNET, an online unsupervised anomaly detector based on an ensemble of auto-encoders that characterizes anomalies by using reconstruction errors.
Zhang et al. \cite{zhang2020dynamic} combined multiple gated CNNs and a bidirectional LSTM to design a deep neural network architecture to process the extracted features.
Similarly, Vasan et al. \cite{vasan2020image} proposed an ensemble convolutional neural network (CNN) based architecture for both packed and unpacked malware detection.

For type (iii), Chen et al. \cite{chen2021adsim} proposed ADSIM, an online, unsupervised, and similarity-aware network anomaly detection algorithm based on heterogeneous ensemble learning.
The MGEL \cite{guo2021mgel} built diverse base learners using multi-grained features and then identified malware encrypted traffic in a stacking way.
Rabadi et al.\cite{rabadi2020advanced} proposed multiple models to study and generalize the heterogeneous API arguments without the need for explicit expertly in the domain.
\textbf{Compared to homogeneous tree ensembles and deep ensembles, heterogeneous ensembles are more relevant for security applications because the appropriate model can be selected based on the different types of features.}

\vspace{-0.15in}
\subsection{Adversarial Attack for Security Detectors}
A more common adversarial attack against security detectors is a \textit{restricted feature-space attacks} - this type of attack obtained a feature vector as input and outputs another perturbed feature vector.
It only needs to ensure that the feature vectors can be transformed into objects in the problem-space by a remapping function.
In NIDSs, Rigaki et al.\cite{rigaki2017adversarial} presented the first try to apply adversarial attacks to NIDSs from the deep learning image classification domain.
Then, Hashemi et al. \cite{hashemi2019towards} ensured that the feature vector can be mapped back to the real traffic by designing various remapping functions.
In the context of malware detection, in order to evade raw bytes-based malware detectors like MalConv \cite{nataraj2011malware}, Kreuk et al. \cite{kreuk2018deceiving} iteratively generated the adversarial payload with the gradient-based method of FGSM in the continuous embedding feature-space.
Similarly, Suciu et al. \cite{suciu2019exploring} extended the FGSM-based adversarial attacks with two previously proposed strategies (i.e., append-FGSM and slack-FGSM).

A small part of the work is also directly employed the \textit{End-to-end problem-space attacks} - this type of attack produced a real sample as an output, so it is a specific real-life attack.
As a classic problem-space attack in the field of NIDSs, Han et al.\cite{han2020practical} first applied a GAN to generate adversarial features, followed by using PSO to make traffic-space vectors close to the adversarial features.
Demetrio et al. \cite{demetrio2020adversarial} proposed a general adversarial attack framework (RAMEn) against PE malware detectors based on two novel functionality-preserving manipulations, namely Extend and Shift, which injected adversarial payloads by extending the DOS header and shifting the content of the first section in PE files.

On the other hand, there has also been some work on ensemble attacks in the security detectors, Al-Dujaili et al. \cite{al2018adversarial} demonstrated the difference between four attacks in a sense that deep neural network-based malware detectors enhanced by training with one attack cannot resist the other attacks.
\cite{li2020adversarial} exploited the “max” strategy by permitting attackers to have multiple generative methods and multiple manipulation sets.

\subsection{Model Ensemble for Defenses}
Several researchers contended that ensemble learning \cite{dietterich2002ensemble} can be used to improve the robustness of classifiers by combining multiple classifiers.
For example, Biggio et al. \cite{biggio2010multiple,biggio2009multiple,biggio2015one} investigated the ways for building a multiple classifier system (MCS) that can improve the robustness of linear classifiers.
In addition, Abbasi et al. \cite{abbasi2017robustness} proposed the “specialists + 1” ensemble, while Xu et al. \cite{xu2017feature} combined various feature squeezing techniques in order to detect AEs.

In the context of security applications,
Shu et al. \cite{shu2022omni} employed the Omni method to create unexpected sets of models whose hyperparameters are controlled to make the attacker's target model far away, for defensive purposes.
Wang et al. \cite{wang2021def} proposed Def-IDS, an ensemble defense mechanism capable of resisting known and unknown adversarial attacks, composed of two parts: a multi-class generative adversarial network and a multi-source adversarial retraining technique.

Nevertheless, some previous studies have not supported the use of ensemble learning for defense purposes.
At first, Zhang et al. \cite{zhang2018gradient} showed that a discrete-valued tree ensemble classifier can be easily evaded by manipulating the model decision outputs only. 
Following that, in a more realistic scenario, they \cite{zhang2020decision} studied evasion attacks on ensemble models without knowing the details of the classifier, in which case ensemble learning is also very easy to avoid.
On the other hand, He et al. \cite{he2017adversarial} demonstrated that adaptive adversaries can easily and successfully create adversarial examples, thus implying that ensemble learning is not sufficient to provide a strong defense against adversarial examples.
\textbf{To conclude, it remains unclear whether ensemble technologies can increase the robustness of a system, so it is imperative to systematically evaluate their robustness in security applications.}

\subsection{Ensemble Adversarial Training}
Defenders against AEs, like adversarial training \cite{goodfellow2014explaining,madry2017towards}, are typically tailored to a single attack type. For other attack methods, these defenses offer no guarantees and, at times, even increase the model’s vulnerability.
Several solutions were proposed by the researchers to solve the problem.
As improved ensemble adversarial training methods, Tramer et al. \cite{tramer2017ensemble} introduced a technique called ensemble adversarial training that can augment training data with perturbations transferred from other models for increasing robustness. 
Kariyappa et al. \cite{tramer2019adversarial} again formulated an ensemble adversarial training and proposed two strategies for ensemble adversarial training, i.e., the "Max" strategy and the "Avg" strategy.
Song et al. \cite{song2018improving} proposed a novel Adversarial Training with Domain Adaptation (ATDA) method.
\textbf{However, the works cited above focus on the \textit{robust generalization} of different types of $L_p$-bounded attacks, rather than the \textit{robust generalization} of different types of attack methods (e.g., FGSM, JSMA and ZOSGD).}

In terms of our most relevant work, \cite{li2020adversarial} exploited the “max” strategy by permitting attackers to have multiple generative methods and multiple manipulation sets.
\textbf{Our work differs somewhat from theirs in three main respects: (i) we extend model ensemble to the entire machine learning domain, explore transfer ensemble attack and additionally use more gradient-free attack methods; (ii) we apply ensemble attacks for stronger adaptive capabilities; (iii) we apply ensemble weight optimization to defend against multiple attack methods as well as adaptive attacks simultaneously}.

\section{preliminaries}

In this section, we first describe the threat model.
Then, we introduce two types of adversarial attacks against security detectors. 
Finally, we summarize the different attack and defense methods used in our framework. 
Table~\ref{table-notations} lists some important notations used in this paper.
Table~\ref{table-abb} lists some Abbreviations and Acronyms for our framework.

\subsection{Threat Model}
We provide here a detailed threat model for adversarial attacks against security detectors.
This threat model includes the definition of the adversary's goals and knowledge of the target system.
\begin{table}[]
    \caption{Notations in Problem Formulation} 
    \label{table-notations}
    \begin{tabular}{@{}ll@{}}
    \toprule
    Notations & Description                                                    \\ \midrule
    $\mathbf{z},\hat{\mathbf{z}}$         & original and adversarial example                               \\
    $\mathcal{Z}$        & problem-space                                                  \\
    $\mathbf{x},\hat{\mathbf{x}}$         & feature vector extracted from original, adversarial example \\
    $\mathcal{X}$         & feature-space                                                  \\
    $\phi,\phi'$         & targeted and substitute feature extractor                      \\
    $f,f'$         & targeted and substitute ML classifier                          \\
    $\mathcal{G}_a,\mathcal{G}_d$         & attack and defense objective function                         \\
    $\mathcal{L}$            & loss function                                                  \\
    $\theta $    & model parameters                                                   \\
    $l$         & distance function between two feature vectors                  \\
    $\mathcal{D}$         & data distribution                                            \\
    $\mathcal{S}$         & perturbation set                                             \\
    $\mathcal{A}$        & attack set                                                    \\
    $\mathbf{w}_a,\mathbf{w}_t$        & attack/adversarial training weight vector                                                    \\
    $\delta$        & perturbation added to the feature vector                       \\
    $\varTheta$         & feature-space constraints                                      \\
    $\varGamma$         & problem-space constraints                                      \\
    $\mathcal{M}$        & remapping function                                            \\ \bottomrule
    \end{tabular}
    \end{table}

\textbf{Adversary's Goals.}
In a security system (e.g., intrusion detection or malware detection), the attackers' goal is to evade the detectors during the testing phase.
Specifically, the attackers might want a specific class ("malicious") classified as another class ("benign").

\textbf{Adversary's Knowledge.}
Based on the prior knowledge of the attacker, there are four types of attacks on ML-based security detectors: transparent-box attacks, white-box attacks, gray-box attacks, and black-box attacks.

\begin{itemize}
    \item Transparent-Box attacks: In this case, the adversary has complete knowledge about the system architecture $f$ and parameters $\theta$, including both white-box knowledge and knowledge about the defense methods used by a defender. Such knowledge can assist the attacker in choosing an adaptive attack that would be capable of bypassing the specific defense mechanism.
    \item White-Box attacks: The attackers know every detail about the machine learning model, including its architecture $f$ and parameters $\theta$, and by calculating the optimization problem based on the ML model, the adversarial examples $\hat{\mathbf{z}}$ can be generated easily.
    \item Gray-Box attacks: Gray-box attacks assume that the attack does not know the specific details of the model, but knows part of the information, such as the feature extractor $\phi$, which can generate an adversarial example by querying the output of the target model $f$.
    \item Black-Box attacks: In a more practical setting, we assume that the attacker does not have any or very limited knowledge about the target model $f$ nor its feature extractor $\phi$.
    In this case, the attacker can only build a substitute model $f'$ and a substitute feature extractor $\phi'$  based on domain knowledge.
\end{itemize}

\begin{table}[]
    \caption{Abbreviations and Acronyms}
    \label{table-abb}
    \begin{tabular}{@{}cccc@{}}
    \toprule
    \toprule
                                                                                &                                                                                      & AEs       & Adversarial Examples                     \\ 
                                                                                &                                                                                      & AT        & Adversarial Training\cite{goodfellow2014explaining,madry2017towards}                     \\
                                                                                &                                                                                      & AE-AG        & Adaptive Ensemble Attack Generation                     \\
 \multirow{2}{*}{\textbf{Terms}}                                               &                                                                                      & AutoED        & Automatic Ensemble Defense                     \\
                                                                                &                                                                                      & EMA        & Ensemble of Multiple Attack                     \\
                                                                                &                                                                                      & TEA        & Transfer Ensemble Attack                     \\
                                                                                &                                                                                      & MA-AT        & Multi-attack AT                     \\
                                                                                &                                                                                      & TE-AT        & Transfer Ensemble AT                     \\ \cmidrule(r){1-4}
                                                                                & \multirow{6}{*}{\begin{tabular}[c]{@{}c@{}}Gradient\\ -based\\ Attacks\end{tabular}} & FGSM      & Fast Gradient Sign Method \cite{goodfellow2014explaining}                \\
                                                                                &                                                                                      & PGD       & Projected Gradient Descent\cite{madry2017towards}               \\
                                                                                &                                                                                      & C\&W      & Carlini and Wagner Attack\cite{carlini2017towards}                \\
                                                                                &                                                                                      & JSMA      & Jacobian-based Saliency Map\cite{papernot2016limitations}       \\
                                                                                &                                                                                      & DF        & DeepFool Attack\cite{moosavi2016deepfool}                          \\
                                                                                &                                                                                      & BIM       & Basic Iterative Method\cite{kurakin2016adversarial}                                      \\ \cmidrule(r){2-4}
    \textbf{Attacks}                                                                              & \multirow{6}{*}{\begin{tabular}[c]{@{}c@{}}Gradient\\ -free\\ Attacks\end{tabular}}  & ZOO       & ZO Optimization\cite{chen2017zoo}                \\
                                                                                &                                                                                      & NES       & Natural Evolutionary Strategies\cite{ilyas2018black}          \\
                                                                                &                                                                                      & ZOSGD     & ZO Stochastic Gradient Descent\cite{ghadimi2013stochastic} \\
                                                                                &                                                                                      & ZOAda     & ZO Adaptive Momentum\cite{chen2019zo}    \\
                                                                                &                                                                                      & BA        & Boundary Attack \cite{brendel2017decision}                          \\
                                                                                &                                               & HSJA      & HopSkipJumpAttack \cite{chen2020hopskipjumpattack}                        \\ \cmidrule(r){2-4}
                                                                                &                      \multirow{3}{*}{\begin{tabular}[c]{@{}c@{}}EMA \&\\ TEA\end{tabular}}                                                                 & Avg-EA    & Average Ensemble Attack\cite{tramer2019adversarial}                     \\
                                                                                &                                                                                      & Max-EA    & Maximum Ensemble Attack\cite{li2020adversarial}                     \\
                                                                                &                                                                                      & Adp-EA    & \textbf{Adaptive Ensemble Attack (Ours)}                   \\\cmidrule(r){1-4}
    \multirow{10}{*}{\textbf{Defenses}}                                                 & \multirow{3}{*}{\begin{tabular}[c]{@{}c@{}}Model\\ Ensemble\end{tabular}}            & TreeEns   & Tree Ensemble\cite{chen2021cost}                            \\
                                                                                &                                                                                      & DeepEns   & Deep Ensemble\cite{shu2022omni,li2020adversarial}                            \\
                                                                                &                                                                                      & HeteroEns & \textbf{Heterogeneous Ensemble (Ours)}                   \\ \cmidrule(r){2-4}
                                                                                &                                                                                      & NAT       & Naive Adversarial Training               \\
                                                                                & \multirow{3}{*}{MA-AT}                                                               & Avg-AT   & Average Ensemble AT \cite{tramer2019adversarial}                                          \\
                                                                                &                                                                                      & Max-AT   & Maximize Ensemble AT \cite{li2020adversarial,tramer2019adversarial}                                        \\
                                                                                &                                                                                      & R-AT   & \textbf{Robust Ensemble AT (Ours)}                                         \\ \cmidrule(r){2-4}
                                                                                & TE-AT                                                                                &    & \textbf{Transfer Ensemble AT (Ours)}                                         \\ \hline
    \multirow{7}{*}{\begin{tabular}[c]{@{}c@{}}\textbf{Utility} \\ \textbf{Metrics}\end{tabular}} & \multirow{3}{*}{Attacks}                                          & $ASR$       & Attack Success Rate                      \\
                                                                                &                                                                                      & $ASR_{avg}$  & Average DSR                                        \\ \cmidrule(r){2-4}
                                                                                & \multirow{4}{*}{Defenses}                                                            & $DSR$       & Defense Success Rate                     \\
                                                                                &                                                                                      & $DSR_{avg}$  & Average DSR                                          \\
                                                                                &                                                                                      & $ODR$       & Original Detection Rate                  \\ \cmidrule(l){1-4} 
                                                                                \toprule
                                                                            \end{tabular}
\vspace{-0.15in}
\end{table}
\vspace{-0.15in}
\subsection{Adversarial Attack for Security Detectors}
As shown in \cite{rosenberg2021adversarial}, in contrast to image-based attacks, the cybersecurity domain can modify feature vectors $\mathbf{x}$ (e.g., flow statistical features) or directly modify source files $\mathbf{z}$ (e.g. PCAP and PE file). 
However, the modification process must maintain the file's executability and maliciousness.
In general, there are two approaches, the first of which is to modify the feature vectors $\mathbf{x}$ directly and then map the feature vectors $\mathbf{x}$ back to the problem-space $\mathcal{Z}$ through a remapping function $\mathcal{M}$. 
The second option is to modify the source file $\mathbf{z}$ directly.
\subsubsection{Restricted Feature-Space attacks}
For a given detector $f$ and a given input object $z$ with feature representation $x \in \mathcal{X} (f(x)=y)$, 
the attacker attempts to minimize the distance $l$ between and $\mathbf{x}$ and $\hat{\mathbf{x}}$ in the feature-space, such that the resulting adversarial examples $\hat{\mathbf{x}} \in \mathcal{X}$ in the feature-space can misclassify the classification models $f$ .
And she uses remapping function $\mathcal{M}(\hat{\mathbf{x}})$ to project $\hat{\mathbf{x}}$ back to legal space for crafting adversarial features $\hat{\mathbf{x}}$, which ensures that $\hat{\mathbf{x}}$ can be transformed into a legal $\hat{\mathbf{z}}$.
We next introduce the remapping function $\mathcal{M}$ for intrusion detection and malware detection in detail.
\begin{itemize}
    \item Network Intrusion Detections (NIDSs): We group flow-based features into the following four groups.
    The first category of features is those we can modify independently.
    The second group of features is dependent on other features;
    Third, some features cannot be changed because attackers cannot control them;
    There is another type of feature in which their value depends on the actual packets of the flow and cannot be calculated by the value of other features;
    The remapping function $\mathcal{M}$ allows us directly modify the features that can be modified independently and calculate the second group of features instantly, without modifying the last two groups of features.
    \item Malware Detections: Consider the two simplest adversarial attacks of malware detection: append and slack attacks.
        The remapping function of malware $\mathcal{M}$ is better defined than network traffic.
        An append attack adds some specially trained bytes at the end of the file to trick the classifier. 
        Slack attacks look for loose pieces of samples (binaries/script files/attack traffic) and insert perturbations in those locations. 
        Thus, the remapping function $\mathcal{M}$ allows only the tail and loose fragments to modify the value, and the rest remains the same.
\end{itemize}
\vspace{-0.1in}
\subsubsection{End-to-end Problem-Space attacks}
In contrast with the aforementioned feature-space attacks, 
the problem-space attacks refer to the adversarial attack that is performed in the problem-space $\mathcal{Z}$, i.e., how to “modify” the real-world input $\mathbf{z} \in \mathcal{Z}$ with a minimal cost (e.g., executables, source code, pcap, etc.) such that the generated AEs $\hat{\mathbf{z}} \in \mathcal{Z}$ in the problem-space can misclassify the target model $f$.

\textbf{As a systematic study of adversarial attacks on security detectors, our framework incorporates existing restricted feature-space attacks and, of course, implements partial end-to-end attacks to ensure that our platform can be extended to problem-space attacks in the future.}

\vspace{-0.1in}
\subsection{Attacks \& Defenses}
In this section, we summarize the state-of-the-art attack and defense methods as shown in Table~\ref{table-abb}.
\vspace{-0.1in}
\subsubsection{Attack Strategies.}

\textbf{Gradient-based attacks.}
We implement 6 gradient-based attacks: Fast Gradient Sign Method (FGSM) \cite{goodfellow2014explaining}, Projected Gradient Descent (PGD) \cite{madry2017towards},
Carlini and Wagner Attack (C\&W) \cite{carlini2017towards},Jacobian-based Saliency Map Attack (JSMA) \cite{papernot2016limitations}, DeepFool Attack (DF) \cite{moosavi2016deepfool} and Basic Iterative Method (BIM) \cite{kurakin2016adversarial}.
At first, Goodfellow et al. \cite{goodfellow2014explaining} proposed a fast method for generating AEs called FGSM.
They perform only one step of gradient update along with the sign direction of the gradient at each pixel.
And then, Madry et al. \cite{madry2017towards} proposed the use of PGD method to solve the linear assumption problem in FGSM \cite{goodfellow2014explaining}.
Following that, a series of relevant works about AEs are provided.
Carlini and Wagner \cite{carlini2017towards} proposed an optimization-based attack method named C\&W;
Papernot et al.\cite{papernot2016limitations} computed the adversarial saliency map to update each input feature;
Moosavi-Dezfooli et al.\cite{moosavi2016deepfool} proposed the DeepFool adversarial method to find the closest distance from the original input to the decision boundary of AEs;
The BIM attack\cite{kurakin2016adversarial} is an iterative version of FGSM where a small perturbation is added in each iteration.

\textbf{Gradient-free attacks.}
Gradient-free is a more practical attack method, and it is more convenient to apply to cybersecurity discrete data.
As a result, we implement 6 gradient-free attack methods, which is more than the general library: Zeroth-order Optimization (ZOO) \cite{chen2017zoo}, Natural Evolutionary Strategies (NES) \cite{ilyas2018black}, 
Zeroth-Order Stochastic Gradient Descent (ZOSGD), Zeroth-Order Adaptive Momentum Method (ZOAda) \cite{chen2019zo}, Boundary Attack (BA) \cite{brendel2017decision} and HopSkipJumpAttack (HSJA) \cite{chen2020hopskipjumpattack}.
At first, Chen et al.\cite{chen2017zoo} presented the zeroth-order optimization (ZOO) attack methods.
Following that, several algorithms are proposed in order to speed up the search.
Ilyas et al. \cite{ilyas2018black} used NES, a method for derivative-free optimization based on the idea of a search distribution;
Ghadimi et al. \cite{ghadimi2013stochastic} aimed to estimate the gradient of the ML model more accurately to generate adversarial feature;
Chen et al. \cite{chen2019zo} utilized adaptive momentum method to further improve search efficiency.
Also, there are some decision-based algorithms proposed.
For searching for the AEs with the smallest distance from the original example, the BA \cite{brendel2017decision} algorithm used a strategy that performs a random walk at the decision boundary of the target detector, looking for a better example constantly.
Chen et al. \cite{chen2020hopskipjumpattack} combined decision-based and zero-order optimization, which is the current best decision-based method.

\textbf{Ensemble attacks.}
We first implement two ensemble adversarial attack methods: \textit{Average ensemble Attack (Avg-EA)} \cite{tramer2017ensemble}, \textit{Maximum ensemble Attack (Max-EA)} \cite{tramer2019adversarial} and then propose ours \textit{Adaptive ensemble Attack (Adp-EA)}.
The three attack methods can be extended to two scenarios: an ensemble of multiple attack methods (EMA) and a transfer ensembles attack (TEA).
In Section~\ref{sec-ae-ag}, these attacks are described in detail.

\subsubsection{Defense Strategies.}
Model ensembles and adversarial training (AT) are the main defense methods of our framework.

\textbf{Model Ensembles.}
Several researchers contended that ensemble learning \cite{dietterich2002ensemble} can be used to improve the robustness of classifiers by combining multiple classifiers.
We implement three types of ensemble models: deep ensembles (\textit{DeepEns}), tree ensembles (\textit{TreeEns}), and heterogeneous ensembles (\textit{HeteroEns}).

\textbf{Adversarial Training.}
As the simplest form of AT, Naive adversarial training (NAT) \cite{goodfellow2014explaining,madry2017towards} is typically tailored to one type of attack, which leaves them vulnerable to other types of adversarial attacks.
We divide AT into two categories based on two attack scenarios: (i) Multi-Attack Adversarial Training (MA-AT), (ii) Transfer Ensemble Adversarial Training (TE-AT).
As for MA-AT, we implement two ensemble adversarial training methods: : \textit{Average Adversarial training (Avg-AT)} \cite{tramer2017ensemble} and \textit{Maximum Adversarial Training (Max-AT)} \cite{tramer2019adversarial}, and propose our \textit{robust ensemble adversarial training (R-AT)}.
In Section~\ref{sec-auto-ed}, these defenses are described in detail.

\begin{figure*}[htbp]
    \centerline{\includegraphics[width=0.99\textwidth]{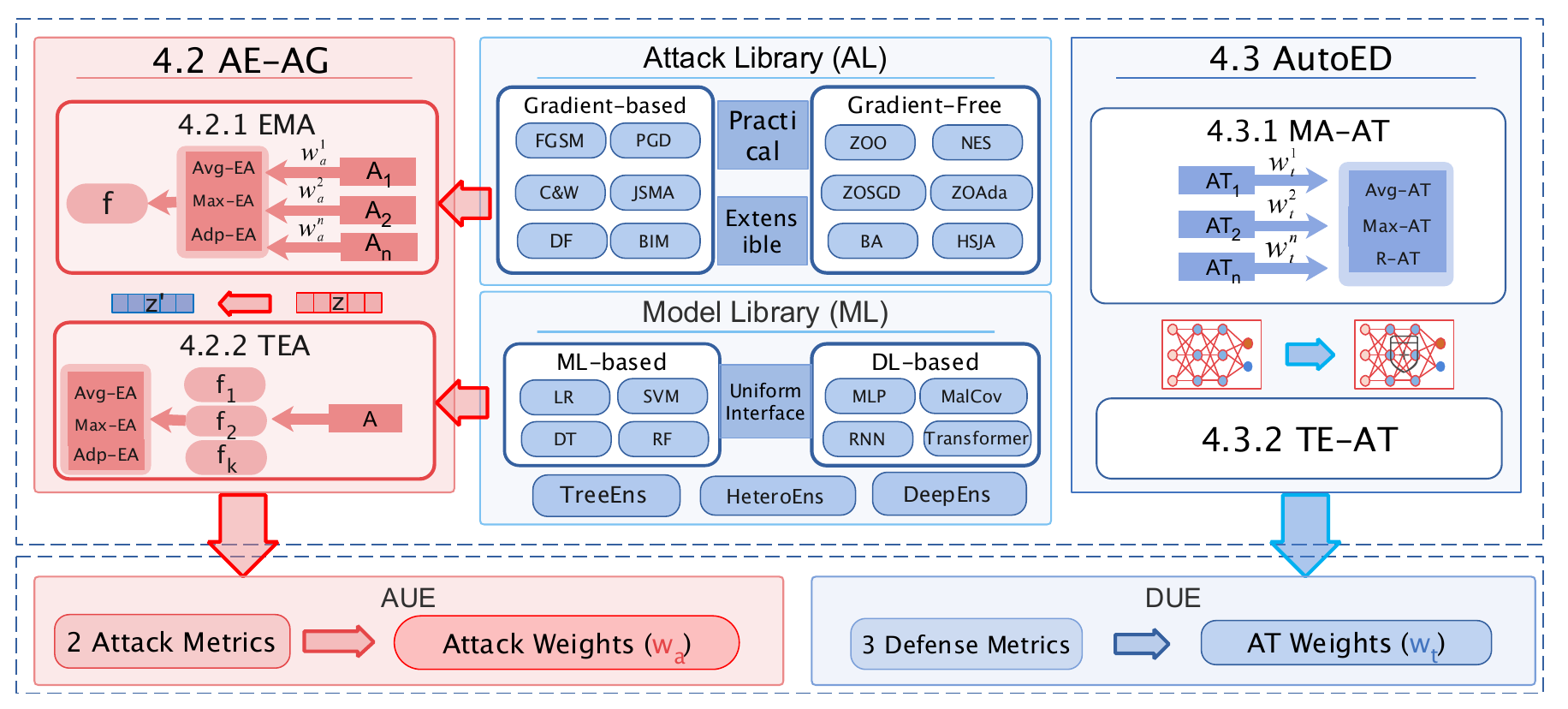}}
    \caption{The CARE framework consists of three basic components and two ensemble robustness evaluation components. The three basic components are: 1) Attack Library (AL) 2) Model Library (ML) 3) Attack and Defense Utility Evaluation (AUE \& DUE); the two ensemble robustness evaluation components are: 1) Adaptive Ensemble Attack Generation (AE-AG) 2) Automatic Ensemble Defense (AutoED).
    }\label{fig_framework}
\end{figure*}

\section{System Design}
\label{sec-system-design}
In this section, we first illustrate the overview of our Cybersecurity Adversarial Robustness Evaluation (CARE) framework.
The CARE framework consists of three basic components and two ensemble robustness evaluation components as shown in Fig.~\ref{fig_framework}.
The two ensemble robustness evaluation components: Adaptive Ensemble Attack Generation (AE-AG) and Automatic Ensemble Defense (AutoED) will be in-depth discussed in Section~\ref{sec-ae-ag} and~\ref{sec-auto-ed}.
\subsection{Overview of CARE}

The Cybersecurity Adversarial Robustness Evaluation (CARE) framework can provide a comprehensive robustness evaluation of model ensembles and ensemble defenses.
First, our framework consists of three basic modules as follows:
\begin{enumerate}
    \item \textbf{Attack Library (AL).} The main function of AL is to exploit vulnerabilities of ML/DL models and attack them via crafting AEs. In this module, we implement 15 state-of-the-art adversarial attacks, including 6 gradient-based attacks, 6 gradient-free attacks, as well as 3 ensemble attacks.
    \item \textbf{Model Library (ML).} The main function of ML is to provide security classifiers. In this module, we implement 12 security classifiers, both traditional machine learning and deep learning models (such as \textit{malimg} and \textit{transformer}), along with three types of ensemble models, deep ensembles (\textit{DeepEns}), tree ensembles (\textit{TreeEns}), and heterogeneous ensembles (\textit{HeteroEns}).
    \item \textbf{Attack and Defense Utility Evaluation (A\&DUE).} AUE implements two utility metrics for adversarial attacks. AUE allows users to assess whether the generated AEs satisfy the essential utility requirements of adversarial attacks; 
    DUE is mainly used to assess the utility of the defenses using three utility metrics. 
\end{enumerate}

In additional, the two ensemble robustness evaluation components are: \textit{Adaptive Ensemble Attack Generation} (AE-AG) and \textit{Automatic Ensemble Defense} (AutoED).

\begin{enumerate}
\item \textbf{Adaptive Ensemble Attack Generation (AE-AG).} In AE-AG, we have mainly measured two adaptive attacks in two cases: \textit{ensemble of multiple attack} methods (EMA) and \textit{transfer ensembles attack} (TEA), which is ensemble of perturbations transferred from other models.
        The attacker will actively probe the detector configuration to adjust his own tactics accordingly.
\item \textbf{Automatic Ensemble Defense (AutoED).} AutoED consists of two parts: 1) \textit{Multi-Attack Adversarial Training} (MA-AT) 2) \textit{Transfer Ensemble Adversarial Training} (TE-AT).
        For MA-AT, we explore a \textit{robust ensemble adversarial defense} (R-AT) method for adaptive attacks.
        Specifically, we propose a unique \textit{robust generalization optimization objective} and use Bayesian optimization to find the most robust ensemble weights, enabling the model to withstand multiple attacks at once, even from adaptive attackers.
\end{enumerate}

\subsection{Adaptive Ensemble Attack Generation (AE-AG)}
\label{sec-ae-ag}
In AE-AG, we have mainly measured two ensemble attack methods: ensemble of multiple attack methods (EMA) and transfer ensembles attack (TEA).
In this section, we will introduce the EMA in detail. 
The attack method for the TEA is similar to that and will not be discussed in detail here.

\subsubsection{Ensemble of multiple attack methods (EMA)}
One of the simplest forms of ensemble attack is the ensemble of multiple attack methods (EMA).
Specifically, the attacker performs a ensemble attack consisting of $n$ different base attacks $\mathcal{A}_1,\mathcal{A}_2,\dots,\mathcal{A}_n$ (e.g., FGSM and JSMA), with attack weight $\mathbf{w}_a=(w_a^1,w_a^2,\cdots,w_a^n$), to craft ensemble AEs $\delta_{ens}$ for attacking the target model $f$. 
Basic attacks can be composed of gradient-based attacks (e.g., FGSM, C\&W) and gradient-free attacks (e.g., ZOO, BA).
Researchers have previously examined a variety of ensemble attacks \cite{li2020adversarial,tramer2019adversarial}, including \textit{Average Ensemble Attack (Avg-EA)} and \textit{Maximum Ensemble Attack (Max-EA) }.
In this paper, we assume that the attacker has a greater ability to adapt its ensemble strategy to the target classifier, called \textit{Adaptive Ensemble Attack (Adp-EA)}.
Accordingly, it can help us train a more robust classifier model.
Although it allows attackers more freedom, it is practical in the context of security applications.
Next, we will formalize these three types of attack as follows:

\begin{itemize}
    \item Avg-EA: The attacker attacks the target classifier $f$ based on the average ensemble strategy $\mathcal{A}_{ens}=\mathcal{A}_{avg}$ of various base attack methods $\mathcal{A}_{1},\mathcal{A}_{2},\dots,\mathcal{A}_{n}$, crafting ensemble perturbations $\delta_{ens}=\frac{1}{n}\sum^n_{i=1}w_a^i\delta_i$, where each attack method $\mathcal{A}_i$ corresponds to a perturbation $\delta_i$. 
    \item Max-EA: The attacker attacks the target classifier $f$ based on the strongest attack strategy $\mathcal{A}_{ens}=\mathcal{A}_{max}$ of various base attack methods $\mathcal{A}_{1},\mathcal{A}_{2},\dots,\mathcal{A}_{n}$, crafting ensemble perturbations $\delta_{ens}=\delta_k$, for $k^{\star}=\arg \max \mathcal{G}_a(\mathcal{A}_k(\mathbf{x}))$.
    \item Adp-EA: In order to maximize the effectiveness $\mathcal{G}_a$ of the attacks, we allow attacker to adjust the attack weight vectors $\mathbf{w}_a=(w_a^1,w_a^2,\dots,w_a^n)$ of his base attack methods $\mathcal{A}_{1},\mathcal{A}_{2},\dots,\mathcal{A}_{n}$ according to the target model $f$. Here $\mathcal{A}_{ens}$ is adaptive ensemble strategy and we apply Bayesian optimization to attack weight optimization in Section~\ref{sec-bayes}.
\end{itemize}

\subsubsection{Transfer Ensemble Attack (TEA)}
While EMA's ensemble domains are different attack methods, TEA's ensemble domains are different base classifiers.
For transfer ensemble attack (TEA), we assume an attacker who has black-box access to the target model, where the attacker does not know the target model’s architecture $\theta $ and does not freely interact with the target models $f$.
Therefore, an attacker has to leverage a transfer-based adversarial attack against an ML-based classifier.
The transfer ensemble attack methods described here can also be divided into three categories: \textit{Average Ensemble Attack (Avg-EA)}, \textit{Maximum Ensemble Attack (Max-EA)} and \textit{Adaptive Ensemble Attack (Adp-EA)}.
We consider adaptive attackers using \textit{Adp-EA} to attack the ML-based detectors based on the transferability of adversarial attacks. 
The substitute classifier $f'$ is an ensemble classifier composed of $k$ base classifier models $f'_1,f'_2,\dots,f'_k$.
Then, the attacker transforms malicious example $\mathbf{x}$ into adversarial example $\hat{\mathbf{x}}$ base on local substitute model $f'$.
Finally, the attacker transfers them to a victim model $f$.

\subsection{Automatic Ensemble Defense}
\label{sec-auto-ed}
We first discuss the two components of AutoED: (i) \textit{Multi-Attack Adversarial Training (MA-AT)},
(ii) \textit{Transfer Ensemble Adversarial Training (TE-AT)}.
Then we model the MA-AT as a minimax game model and finally present a \textit{robust ensemble adversarial training (R-AT)} method for simultaneously resisting multiple attack methods, along with adaptive attacks.

\subsubsection{Multi-Attack Adversarial Training (MA-AT)}
\label{sec-ma-at}

Conventional adversarial training (AT) is restricted to a single type adversarial attack \cite{goodfellow2014explaining,madry2017towards} as folows:
\begin{equation}
    \min _{\theta} \mathbb{E}_{(x, y) \sim \mathcal{D}}\left[\max _{\delta \in \mathcal{S}} \mathcal{L}(\theta, x+\delta, y)\right]
    \end{equation}
However, there possibly exist blind attacking spots across multiple types of adversarial attacks so that AT under one attack would not be strong enough against another attack.
Thus, an interesting question is how to generalize AT under multiple types of adversarial attacks.
One possible way is to use the finite-sum formulation in the adversarial training inner maximization problem.
AT can be generalized against the strongest adversarial attack across $n$ attack types to avoid blind attacking spots.
There are two strategies for multi-attack adversarial training: \textit{Average Adversarial training (Avg-AT)} \cite{tramer2017ensemble} and \textit{Maximum Adversarial Training (Max-AT)} \cite{tramer2019adversarial}.
\begin{itemize}
       \item Max-AT: For input $\mathbf{x}$, we train on the strongest adversarial example from all attacks, i.e., the max in $\mathcal{L}$ is replaced by $L\left(f\left(\mathcal{A}_{k^{*}}(\mathbf{x})\right), y\right)$, for $k^{*}=\arg \max _{k} L\left(f\left(\mathcal{A}_{k}(\mathbf{x})\right), y\right)$.
       \item Avg-AT: This strategy simultaneously trains on adversarial examples from all attacks. 
        That is, the max in $\mathcal{L}$ is replaced by $\frac{1}{n} \sum_{i=1}^{n} L\left(f\left(\mathcal{A}_{i}(\boldsymbol{x}), y\right)\right)$.
\end{itemize}

\subsubsection{Transfer Ensemble Adversarial Training (TE-AT)}
The previous research \cite{tramer2017ensemble}  has shown that the model thus learns to generate weak perturbations, rather than defend against strong ones. As a result, we find that adversarial training remains vulnerable to black-box attacks, where we transfer perturbations computed on undefended models. 
We support ensemble adversarial training methods against such transfer attacks, but further experiments show that previous ensemble adversarial training methods are sufficient to defend against transfer ensemble attacks, so such adversarial training is not the topic of the study.

\subsubsection{Minimax Game for Automatic Ensemble Defense}
The two multi-attack adversarial training methods in Section~\ref{sec-ma-at} cannot withstand the adaptive attack, as we know, game theory is usually used to tackle adaptability attacks of such an attacker.
Thus, we use \textit{minimax game} as a more general approach to robust generalization of AT.
We formulate MA-AT as a \textit{Minimax Game} defined in Equation~\ref{equ-game}, where the attackers will vary their ensemble attack strategies $\mathbf{w}_a$ to maximize the attack effect $\mathcal{G}_a=\sum_{i=1}^n w_i \mathcal{L}(\mathcal{A}_i(x),y)$. 
The ensemble attack strategies $\mathcal{A}_{ens}$ is composed of base attacks $\mathcal{A}_1,\mathcal{A}_2,\dots,\mathcal{A}_n$ with different weights $\mathbf{w_a}=(w_a^1,w_a^2,\dots,w_a^n)$.
And then attackers sample ensemble AEs $\delta_{ens}$ according to the current ensemble strategy $\mathcal{A}_{ens}$.
At the same time, the detector will reconfigure its detection strategy, which is represented by adversarial training weight $\mathbf{w_t}=(w_t^1,w_t^2,\dots,w_t^n)$ for different attacks methods to minimize the updated $\mathcal{G}_a=\sum_{i=1}^n w_i \mathcal{L}(\mathcal{A}_i(x),y)$. 

Minimax Game for Automatic Ensemble Defense.
Let
\begin{equation}
\label{equ-game}
\min_{\theta} \mathbb{E}_{(x, y) \sim \mathcal{D}} \max_{\mathbf{w}\in\mathcal{W},\delta_i \in \mathcal{S}_i} \sum_{i=1}^n w_i \mathcal{L}(\mathcal{A}_i(x),y)
\end{equation}
where $\mathcal{L}$ is the loss function and $\mathcal{D}$ is the training set.
For $i\in [1,n]$, $\mathcal{A}_i$ is different attacks and $n$ is the number of attack type.

However, adversarial optimization for attackers and defenders is computationally expensive, so we next consider a more lightweight approach.

\subsubsection{Robust Ensemble Adversarial Training}
In order to achieve robust generalization capability of adversarial training, we \textit{propose Robust Ensemble Adversarial Training (R-AT)}.
Our goal is to achieve robustness in the context of adaptive evolutionary attackers by simultaneously focusing on different types of mixed strategies.
First, we define new \textit{robust generalization optimization objectives}:

\textbf{Robust Generalization Optimization Objectives.}
If you want to resist multiple attacks at once, optimizing the detection rate of a single type attack will not be enough.
We are confident that a security detector after AT can resist AEs that are generated by multiple attack methods simultaneously, it will also be able to resist the mixture of multiple attack methods.
For MA-AT, we need to optimize $\mathbf{w}_t=(w_t^1,w_t^2,\dots,w_t^n)$.
In order to determine the most robust weight vectors $\mathbf{w}_t^{\star}$, we develop a new optimization objective that is resistant to AEs generated by various attack methods simultaneously, which we call all AEs defense success rate $DSR_{all}$.
Specifically, for a original sample $\mathbf{x}$, we employ multiple attack methods $\mathcal{A}_1,\mathcal{A}_2,\dots,\mathcal{A}_n$ to generate multiple AEs $\hat{\mathbf{x}}_1,\hat{\mathbf{x}}_2,\dots,\hat{\mathbf{x}}_n$.
And each samples $\mathbf{x}_1,\mathbf{x}_2,\dots,\mathbf{x}_c$ is labeled $\mathbf{y}=[y_1,y_2,\dots,y_c]$, the outputs of all AEs that input to the model $f$ are $\mathbf{y}^a_1,\mathbf{y}^a_2,\dots,\mathbf{y}^a_n$.
We define $\mathbf{b}=[b_1,b_2,\dots,b_c]$ is bool vectors, and $b_i=True \ (y_i==y^a_i)$, $\mathbf{b}^a_1,\mathbf{b}^a_2,\dots,\mathbf{b}^a_n$ corresponds to $\mathbf{y}^a_1,\mathbf{y}^a_2,\dots,\mathbf{y}^a_n$.
The $DSR_{all}$ is defined as follows:
\begin{equation}
    DSR_{all}:=sum_{true}(\mathbf{b}^a_1\&\&\mathbf{b}^a_2,\dots,\&\&\mathbf{b}^a_n)/len(\mathbf{y})
\end{equation}
where $sum_{true}$ is a function that represents the number of $True$ in vector $\mathbf{b}$, $\&\&$ denotes the logical AND operation, $len$ denotes the number of elements in vector $\mathbf{y}$.

Thus, given a space of adversarial training weight vectors $\mathcal{W}$, weight optimization can be represented as follows:
\begin{equation}
    \mathbf{w}_t^{\star}=\underset{ \mathbf{w}_t \in \mathcal{W}}{\operatorname{argmax}} \mathcal{G}_d
\end{equation}
Here $\mathcal{G}_d$ represents the objective to be maximized, here it is $DSR_{all}$. 
$\mathbf{w}_t^{\star}$ is the weight vector that produces the best score, while $\mathbf{w}_t$ can be any value from domain $\mathcal{W}$.
Following that, we solve for the optimal adversarial training weight vector $\mathbf{w}_t$ using Bayesian optimization.

\subsection{Bayesian Optimization Algorithm}
\label{sec-bayes}
For both \textit{adaptive attack (Adp-EA)} and \textit{multi-attack adversarial training (MA-AT)}, the optimal weight needs to be solved via Bayesian optimization. 
They only differ in their optimization goals, but the process is the same.
Therefore, here we are using \textit{multi-attack adversarial training (MA-AT)} as an example to demonstrate the entire optimization process.
Specifically, we employ the Bayesian Optimization (BayesOpt) algorithm for finding the best solution $\mathbf{w}_t^*$.
The $\mathbf{w}_t^*$ is the result of robust ensemble adversarial training (R-AT). 

\noindent\textbf{The reasons for choosing BayesOpt.}
Discrete optimization algorithms include policy gradient, swarm intelligent optimization, and Bayesian optimization.
The policy gradient easily converges to a locally optimal solution.
Additionally, the convergence rate of swarm intelligence optimization is too slow and requires too many attempts.
Adp-EA, especially MA-AT, are expensive per attempt, meaning the swarm intelligence optimization algorithm is not suitable.
Our Bayesian approach differs from the two methods above.
It saves time overhead by referring to previous evaluations when trying the next set of hyper-parameters and can get better results with a small number of attempts. 

\noindent\textbf{Algorithm.}
Bayesian optimization is an approximate optimization method that utilizes a surrogate function to fit the relationship between weight parameters and model evaluation, then selects promising weight parameters for iteration, and ultimately retrieves the best weight parameters.
A Bayesian optimization problem generally consists of four components:
\begin{itemize}
    \item Objective function $\mathcal{G}$: $\mathcal{G}$ is the objective function we're trying to minimize or maximize. For MA-AT, the $\mathcal{G}$ is $DSR_{all}$.
    \item Weight parameters search space $\mathcal{S}$: $\mathcal{S}$ is the weight parameter range to be searched.
    \item Acquisition function $\mathcal{U}$: $\mathcal{U}$ is the function to select the appropriate data $\mathcal{W}_t=(\mathbf{w}_t^1,\mathbf{w}_t^2,\dots,\mathbf{w}_t^k)$. In this paper, we use Gaussian Process-Upper confidence bound (GP-UCB) \cite{srinivas2009gaussian} function.
          The Acquisition function seeks to make each sample as close as possible to the maximum/minimum value of the target function $\mathcal{G}$, in order to increase the efficiency of searching for extreme value points.
    \item Surrogate function $\mathcal{P}$: $\mathcal{P}$ is the function that is fitted to the dataset $\mathcal{W}_t$. We adopt Gaussian Process (GP) in this paper. 
\end{itemize}
\begin{algorithm}[htb]  
	\caption{Algorithm for solving approximate robust weights}  
    \label{alg-bayes}
	\hspace*{0.02in} {\bf Input:} 
    $\mathcal{G}$,$\mathcal{S}$,$\mathcal{P}$\\
    \hspace*{0.02in} {\bf Output:} $\mathbf{w}_t^*$
    \begin{algorithmic}[1] 
        \State Initialize $\mathcal{G}$,$\mathcal{S}$,$\mathcal{P}$
        \State Randomly choose dataset $\mathcal{W}_t=(\mathbf{w}_t^1,\mathbf{w}_t^2,\dots,\mathbf{w}_t^k)$
            \While{: each $ i=1,2,3,...$ }
                    \State $\mathcal{P}(\mathbf{w}_t)=fit(\mathcal{W}_t,g_i)$
                    \State $\mathbf{w}_t^i=\arg\max_{\mathbf{w}_t^i\in \mathcal{S} } \mathcal{P}(\mathcal{W}_t,g_i)$
                    \State $\mathcal{W}_t=\mathcal{W}_t\bigcup (\mathbf{w}_t^i,g_i)$
                    \State Increment $i$
            \EndWhile
            \State  Defender update strategies $\mathbf{w}_t^*$ according to $\mathcal{P}(\mathbf{w}_t)$ 		 
	\end{algorithmic}  
\end{algorithm}

Following that, we describe the Bayesian optimization process.
In each iteration of BayesOpt $i \in \left\{1,\dots,k\right\}$, the defenders first randomly choose dataset $\mathcal{W}_t=(\mathbf{w}_t^1,\mathbf{w}_t^2,\dots,\mathbf{w}_t^k)$ to train the $\mathcal{P}$ and get the target values $g_i$ based on the acquisition function $\mathcal{U}$.
Then, they uses $\mathcal{P}$ to fit $(\mathcal{W}_t,g_i)$, uses the acquisition function $\mathcal{U}$ to select the best $\mathbf{w}_t^*$.
Finally, they add $\mathbf{w}_t^*$ into the model, get a new $g_i$, and re-enter the loop.
Experiments show that BayesOpt can always find an optimal adversarial training weight  $\mathbf{w}_t^\star$ regardless of what $\mathbf{w}_a$ is.
Detail of the proposed BayesOpt Algorithm is described in Algorithm~\ref{alg-bayes}.

\section{experiment Evaluation}

In this section, we first empirically evaluate the effectiveness of 15 adversarial attacks and 8 adversarial defenses on five security datasets against 12 machine learning models.
Second, we verify that simple model ensembles and adversarial training are not resistant to adaptive attackers.
We finally measure the effectiveness of our \textit{multi-attack adversarial training (MA-AT)} and \textit{transfer ensemble adversarial training (TE-AT)}, respectively.
Overall, our experiments cover the following aspects:
\begin{itemize}
    \item \textbf{Experimental Setup.} We introduce implementation and setup of ML-based security detectors, datasets, evaluation metrics in this study. (\ref{sub-ex-es})
    \item \textbf{Basic Evaluation of Attacks.} We test the effectiveness of 12 basic and 3 ensemble adversarial attack methods on four security datasets. (\ref{sub-ex-attack})
    \item \textbf{Basic Evaluation of Defense.} We compare and evaluate basic adversarial defense methods, including three model ensembles and naive adversarial training (NAT) methods. (\ref{sec-ex-def})
    \item \textbf{Evaluation of Adaptive Attacks.} We evaluate adaptive attacks in two scenarios: (1) ensembles of multiple attacks (EMT), and (2) transfer ensemble attacks (TEA). (\ref{sec-adaptive-attacks})
    \item \textbf{Multi-Attack Adversarial Training.} Comparing \textit{multi-attack adversarial training (MA-AT)} methods, we demonstrate that only our \textit{robust ensemble adversarial training (R-AT)} method can withstand multiple attack methods and adaptive attacks simultaneously. (\ref{sec-ex-ma-at}) 
    \item \textbf{Transfer Ensemble Adversarial Training.} We test whether general \textit{transfer ensemble adversarial training (TE-AT)} can withstand multiple transfer attacks simultaneously. (\ref{sec-ex-teat})
\end{itemize}
\begin{table}[]
    \caption{THE CHARACTERISTICS OF DATASETS DURING TRAINING AND TESTING PHASE}
    \label{table-datasets}
\begin{center}

    \begin{tabular}{@{}ccccc@{}}
    \toprule
    \multirow{2}{*}{Data type} & \multirow{2}{*}{Datesets} & \multirow{2}{*}{Feature} & \multicolumn{2}{c}{Samples} \\ \cmidrule(l){4-5} 
                               &                           &                          & Benign      & Malicious     \\ \cmidrule(r){1-3}
    NIDS                       & CICIDS2017                & 80                       & 11632       & 7744          \\
    Malware                    & Bodmas                    & 100                      & 77142       & 57293         \\
    Spam                       & TwitterSpam               & 25                      & 236146      & 186526        \\
    EMTD                       & CICAndMal2017             & 80                       & 6647        & 5516          \\ 
    PE Malware                 & PEData                    & $2^{20}$                      & 2000       & 2000         \\\bottomrule
    \end{tabular}
\end{center}
    \vspace{-0.2in}
\end{table}

\subsection{Experimental Settings}
\label{sub-ex-es}
Here we discuss the settings of our experiments.

\subsubsection{Hardware $\&$ Software}
We first provide statements of hardware and software in our study.

\begin{itemize}
    \item \noindent\textbf{Hardware.} Our experiments are evaluated on a MacBook Pro using 64-bit MacOS system with Intel Core i7-9700K 3.6 GHz 8-Core Processor, 16 GB physical memory, 256GB SSD.
    \item \noindent\textbf{Software.} The software implementation of NashAE is based on Python 3 with several packages, including scikit-learn for ML models and pytorch for DL models.
\end{itemize}

\subsubsection{Dataset}
We select five datasets as shown in Table~\ref{table-datasets}.
Meanwhile, we divide the sampled data into three parts according to 3:1:1 for model training, verification and testing respectively.
The data for training and testing are shown in Table~\ref{table-datasets}, and each part contains both benign and malicious samples. 
The first four datasets are available and widely used in the security literature. 
The four datasets include a network intrusion detection dataset (CICIDS2017), a malware dataset (Bodmas), a  encryption malware traffic detection (EMTD) dataset (CICAndMal2017), and a spam detection dataset (TwitterSpam).
We have collected samples of Windows malware in our last dataset.
\begin{table*}[]
    \begin{center}
      \caption{The effectiveness of attack methods was compared using different attack methods, including gradient-based attacks and gradient-free attacks for CICIDS2017 dataset.}
      \label{attack-measure}
    \begin{tabular}{@{}lllccclclclclccl@{}}
    \toprule
    \multirow{2}{*}{Type} & \multirow{2}{*}{ML-Classifier} & \multicolumn{2}{c}{Detection} & \multicolumn{6}{c}{Gradient-based Attack (\%)} & \multicolumn{6}{c}{Gradient-free Attack (\%)} \\ \cmidrule(l){3-16} 
                          &                                & F1            & R             & FGSM   & PGD    & C\&W     & JSMA  & DF & BIM     & ZOO   & NES   & ZOSGD & ZOAda & BA    & HSJA  \\ \cmidrule(r){1-2}
    Deep                  & MLP                            & 0.91          & 0.98          & \textbf{99.48}  & \textbf{99.48}  & 97.67  & 98.71 & 97,24      & 98.48 & 2.32  & 65.63 & 84.75 & 85.97 & 65.12 & 76.56 \\
                          & CNN                            & 0.92          & 0.97          & 93.48  & 92.38  & 97.67  & 98.71 & 97.45    & \textbf{98.48} & 3.32  & 45.63 & 64.75 & 89.48 & 63.23 & 78.45 \\
                          & RNN                            & 0.93          & 0.98          & 95.48  & 93.48  & 95.67  & 95.45 & \textbf{98.24}      & 97.28 & 2.32  & 55.63 & 82.75 & 88.23 & 62.35 & 69.12 \\
                          & Transformer                    & 0.86          & 0.98          & 97.48  & 96.48  & 93.67  & 96.71 & 93.23      & \textbf{98.48} & 2.32  & 55.63 & 82.75 & 88.56 & 65.67 & 68.35 \\
    Gradient              & LR                             & 0.91          & 0.98          & 6.20   & 6.20   & 6.20   & \textbf{98.44} & \textbf{98.44}    & 6.20    & 3.35  & 5.94  & 5.94  & 5.94  & 4.39  & 39.79 \\
                          & SVM                            & 0.93          & 0.98          & 97.93  & 97.93  & 1.55   & 98.44 & 98.44    & 97.93   & 2.58  & 2.06  & 2.06  & 2.06  & 6.46  & \textbf{99.74} \\
    Tree                  & DT                             & 0.98          & 0.99          & N/A    & N/A    & N/A    & N/A   & N/A      & N/A     & 99.74 & 33.85 & 33.85 & 62.27 & \textbf{99.74}   & \textbf{99.74} \\
                          & RF                             & 0.97          & 0.98          & N/A    & N/A    & N/A    & N/A   & N/A      & N/A     & 13.69 & 37.72 & 37.72 & 37.72 & \textbf{97.93}   & \textbf{97.93} \\
                          & Xgboost                        & 0.98          & 0.99          & N/A    & N/A    & N/A    & N/A   & N/A      & N/A     & 21.96 & 79.84 & 72.60 & 63.30 & 87.86 & \textbf{97.67} \\
    Ensemble              & DeepEns                        & 0.89          & 0.98          & \textbf{98.45}  & \textbf{98.45}  & 67.23  & 58.76 & 95.87    & 57.89   & 39.20  & 54.01 & 63.31 & 56.85 & 57.62 & 65.61 \\
                          & TreeEns                        & 0.98          & 0.99          & N/A    & N/A    & N/A    & N/A   & N/A      & N/A     & 50.85 & 56.35 & 56.35 & 63.31 & \textbf{99.74} & \textbf{99.74} \\
                          & HeteroEns                      & 0.98          & 0.99          & N/A    & N/A    & N/A    & N/A   & N/A      & N/A     & 49.75  & 68.99 & 62.53 & 62.53 & \textbf{96.64} & \textbf{96.64} \\ \bottomrule
    \end{tabular}
  \end{center}
    \end{table*}
    \begin{table}[]
        \begin{center}
        \caption{The default ML used in our experiment.} \label{Table-ML}
        \begin{tabular}{ccc}
        \hline
        Type                      & ML Technique & Differentiable \\ \hline
                        & MLP          & Yes            \\
        Deep Model                          & CNN (MalCov) & Yes            \\
                                  & RNN          & Yes            \\
                                  & Transformer  & Yes            \\ \hline
        Gradient ML               & LR           & Yes            \\
                                  & SVM          & Yes            \\\hline
                              & DT          & No             \\
        Tree                        & RF          & No             \\
                                  & Xgb.         & No             \\ \hline
        \multirow{3}{*}{Ensemble} & DeepEns      & Yes            \\
                                  & TreeEns      & No             \\
                                  & HeteroEns    & No             \\ \cline{1-3} 
        \end{tabular}
        \end{center}
        \end{table}

\noindent\textbf{CIC-IDS2017} 
\cite{sharafaldin2018toward} collects traffic for common attacks in a large-scale testbed, covering many common devices and middleboxes in the network.

\noindent\textbf{CICAndMal2017.}
CICAndMal2017 \cite{lashkari2018toward} is an Android malware dataset that collects 426 malicious and 1,700 benign apps collected by researchers at the University of New Brunswick (UNB) from 2015 to 2017. 
There were four types of malicious samples (adware, ransomware, scareware, and SMS malware) and 42 families. 
During installation, before and after reboot, each malicious sample was run on an actual Android smartphone and network traffic was captured.

\noindent\textbf{Bodmas}
The Bodmas dataset \cite{yang2021bodmas} contains 57,293 malware samples and 77,142 benign samples (134,435 in total).
From August 29, 2019 to September 30, 2020 (one year), malware is randomly selected from a security company's internal malware database.

\noindent\textbf{TwitterSpam.}
The dataset used in Kwon \cite{kwon2017domain} et al.'s research to detect spam URLs on Twitter is publicly available.
Spammers use social networks like Twitter to distribute malware, scams, and phishing content.
Eventually, these URLs reach a landing page containing harmful content after several redirects.

\noindent\textbf{PEData.}
A scalable end-to-end problem-space attack was tested using 2000 PE malware samples and 2000 benign samples.
The malware samples were gathered from VirusTotal5, while the benign samples were downloaded from GitHub.

\subsubsection{ML classifiers.}
We explore gradient models, such as multi-layer perception machine (MLP), recurrent neural networks (RNN), convolutional neural networks (CNN), logistic regression (LR), support vector machines (SVM), decision trees (DT), random forest (RF), XGboost (Xgb.) as tree models, and three typical ensembles: Tree Ensemble (TreeEns), Deep Ensemble (DeepEns) and Heterogeneous Ensemble (HeteroEns).
Furthermore, we examine Transformer and MalConv, a revamped version of CNN in the malware domain.
The baseline ML classifiers are shown in Table ~\ref{Table-ML}.

\noindent\textbf{Transformer.}
To overcome both RNN and encoder-decoder systems' limitations,
This algorithm replaces RNN with attention mechanisms in the seq2seq encoder-decoder.
Transformers have a very long-term memory due to the attention mechanism.
It is possible for a Transformer model to "attend" or "focus" on all of the previous tokens. 

\noindent\textbf{MalConv.}
The MalConv is an end-to-end convolutional neural network (CNN) proposed by Raff \cite{nataraj2011malware} et al.
Taking the first 2MB of an executable as input, it determines if it is malware.
If the input executable length exceeds this threshold, the file is truncated to the specified size, otherwise, it is padded with the value 0.

\subsubsection{Evaluation Metrics}
Metrics used in this work can be classified into two categories:
\begin{itemize}
    \item \noindent\textbf{Detection Performance}: For performance evaluation, we use typical machine learning metrics, including Precision (P), Recall (R), F1-score (F1) and ROC/AUC (AUC).
    \item \noindent\textbf{Attack Effectiveness ($ASR$ and $ASR_{avg}$)}: Specifically, Attack Success Rate ($ASR$) measures the percentage of AEs that have already evaded the detectors. 
                          ($ASR_{avg}$) is a metric to evaluate the average attack success rate against all models.
    \item \noindent\textbf{Defense Effectiveness ($DSR$ and $DSR_{avg}$} Defense Success Rate ($DSR$) indicates that the percentage of AEs that do not successfully attack the detector.
                                                  $DSR_{avg}$ is a metric to evaluate the average defense success rate against all models.
    \item \noindent\textbf{Original Detection Rate ($ODR$).} The OSR is the detection rate of the model after adversarial training on the original data.
\end{itemize}

\subsection{Basic Evaluation of Attacks}
\label{sub-ex-attack}
In this section, we first evaluate the effectiveness of different adversarial attacks on different models, 
then we analyze the effect of different attack costs on the effectiveness of adversarial attacks, and finally, we test an end-to-end problem-space attack in malware detection to demonstrate that our framework can be extended to real-world attacks.

\subsubsection{Attack Effectiveness Comparison (Different Attack Methods)}
First, we compare the effectiveness of 6 gradient-based attacks and 6 gradient-free attacks.
To be fair, for attacks such as JSMA, DeepFool, BA, and HSJA, which are biased towards modifying fewer features, we use $l_1$ norm
to limit the attack cost ($\epsilon $=1), while for other attacks, we use $l_\infty$ to limit the attack cost ($\epsilon $=0.05).
As shown in Table \ref{attack-measure}, gradient-based attacks have a higher success rate than gradient-free attacks at the same cost.
The strongest gradient-free attack, ZOAda, achieves only 85.97\% ASR on MLP, but 99.48\% using FGSM.
Other models show almost identical results.
All gradient-based attacks do not explicitly show the strongest attack effectiveness, with different results for different situations. 
In contrast, HSJA shows excellent performance in models that cannot be solved with gradients, with an average of over 95\% ASR.
In addition, ZOAda with added momentum has the stronger attack effect, with an average 10\% increase over ZOSGD.
\begin{figure*}
  \centering
  \begin{subfigure}{.49\textwidth}
  \includegraphics[width=\textwidth]{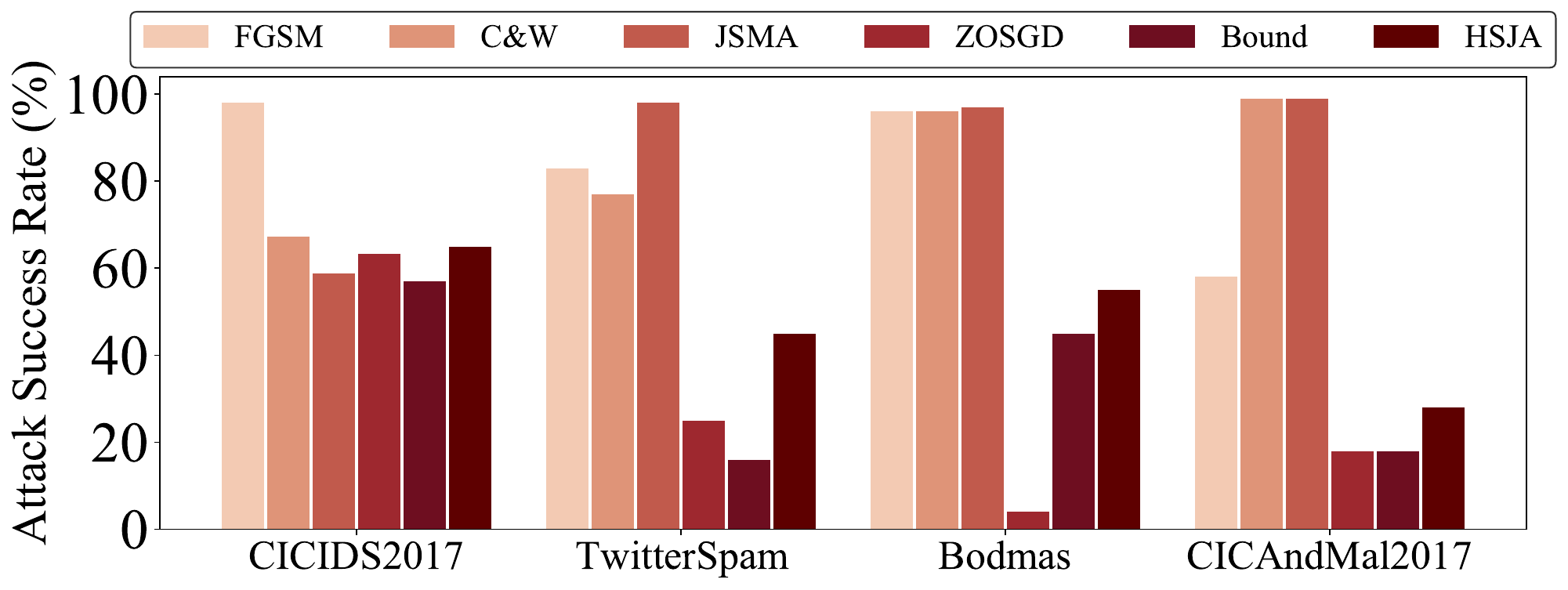}
  \caption{Deep Ensemble}
  \end{subfigure}
  \hfill
  \begin{subfigure}{.49\textwidth}
  \includegraphics[width=\textwidth]{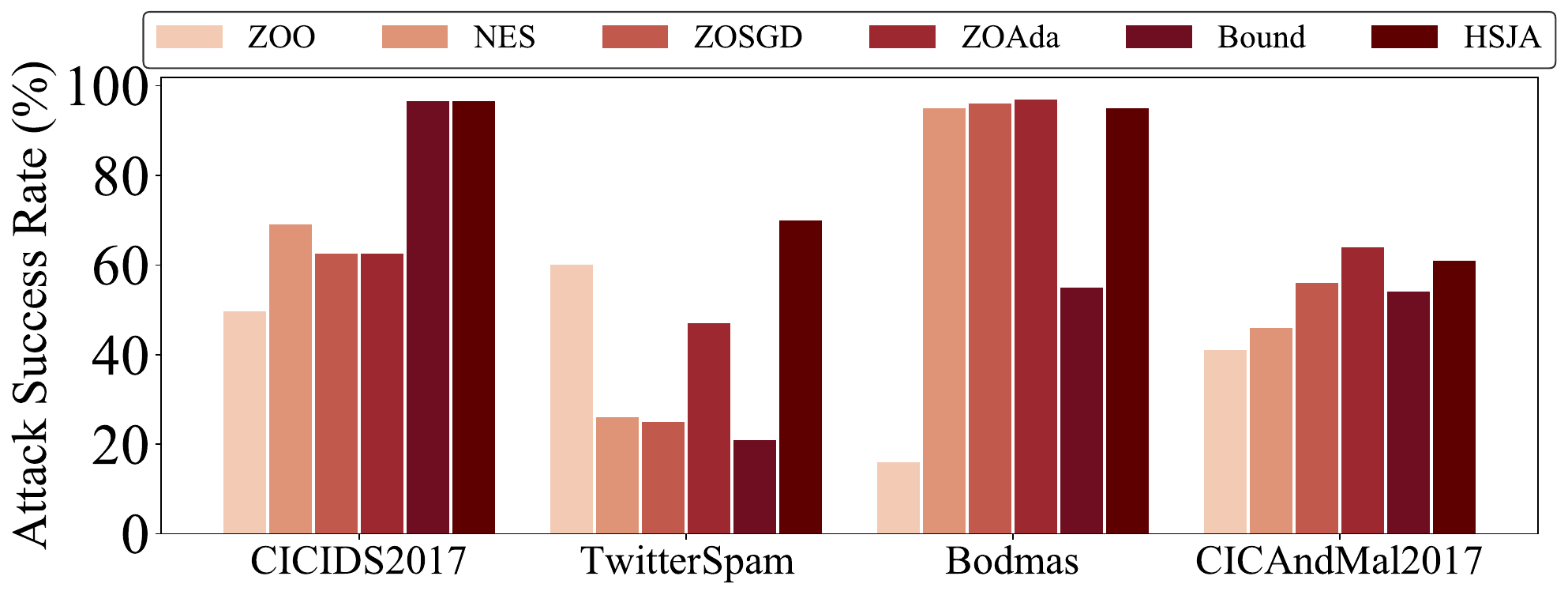}
  \caption{Heterogeneous Ensemble}
  \end{subfigure}
  \caption{Adversarial attack effectiveness on different datasets (\emph{higher is better}).}
  \label{fig-datasets}
  \vspace{-0.06in}
\end{figure*}
\begin{figure*}
  \begin{minipage}[b]{0.5\textwidth} 
  \begin{subfigure}{0.5\textwidth}
   \includegraphics[width=\textwidth]{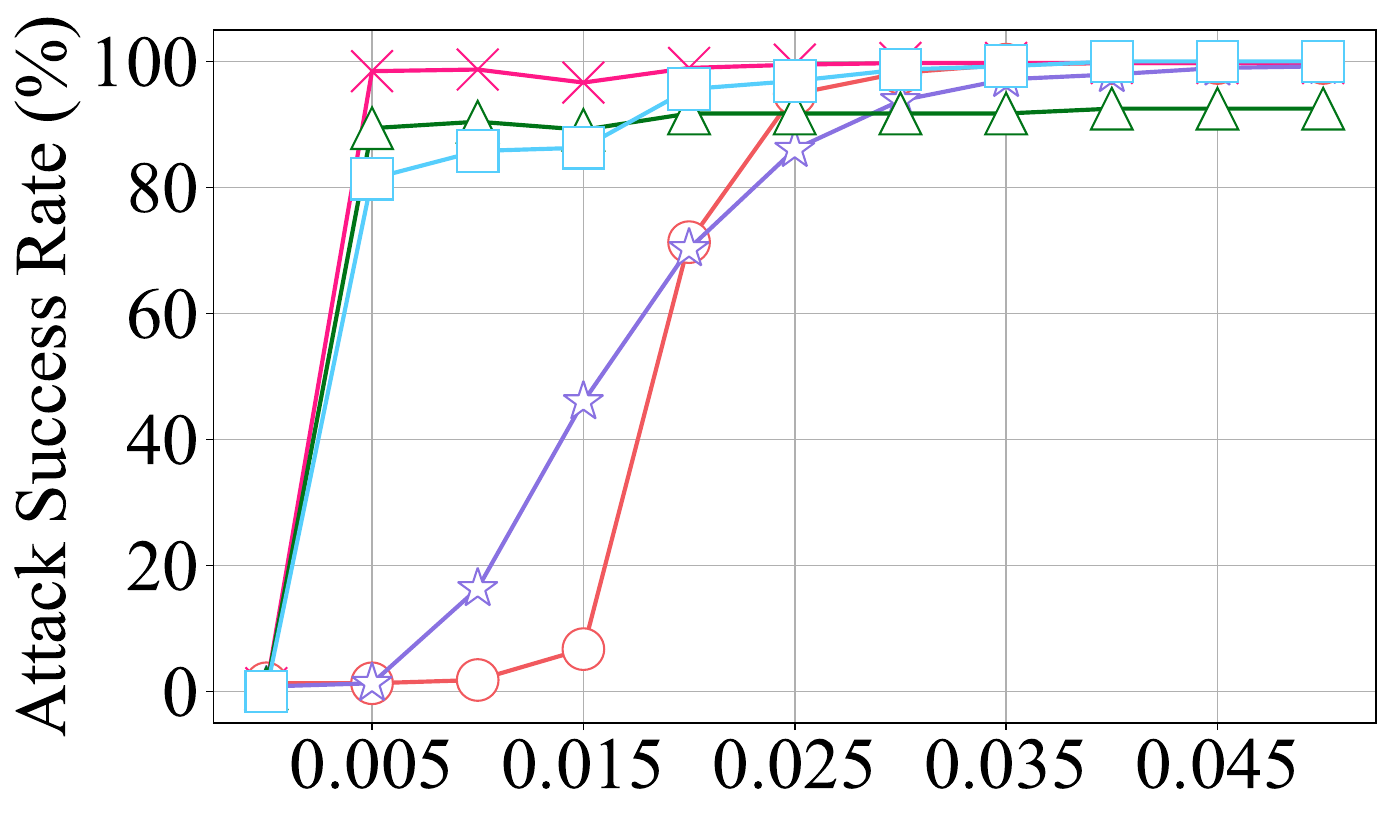}
   \caption{CICIDS2017}
  \end{subfigure}
  \hfill
  \begin{subfigure}{0.5\textwidth}
    \includegraphics[width=\textwidth]{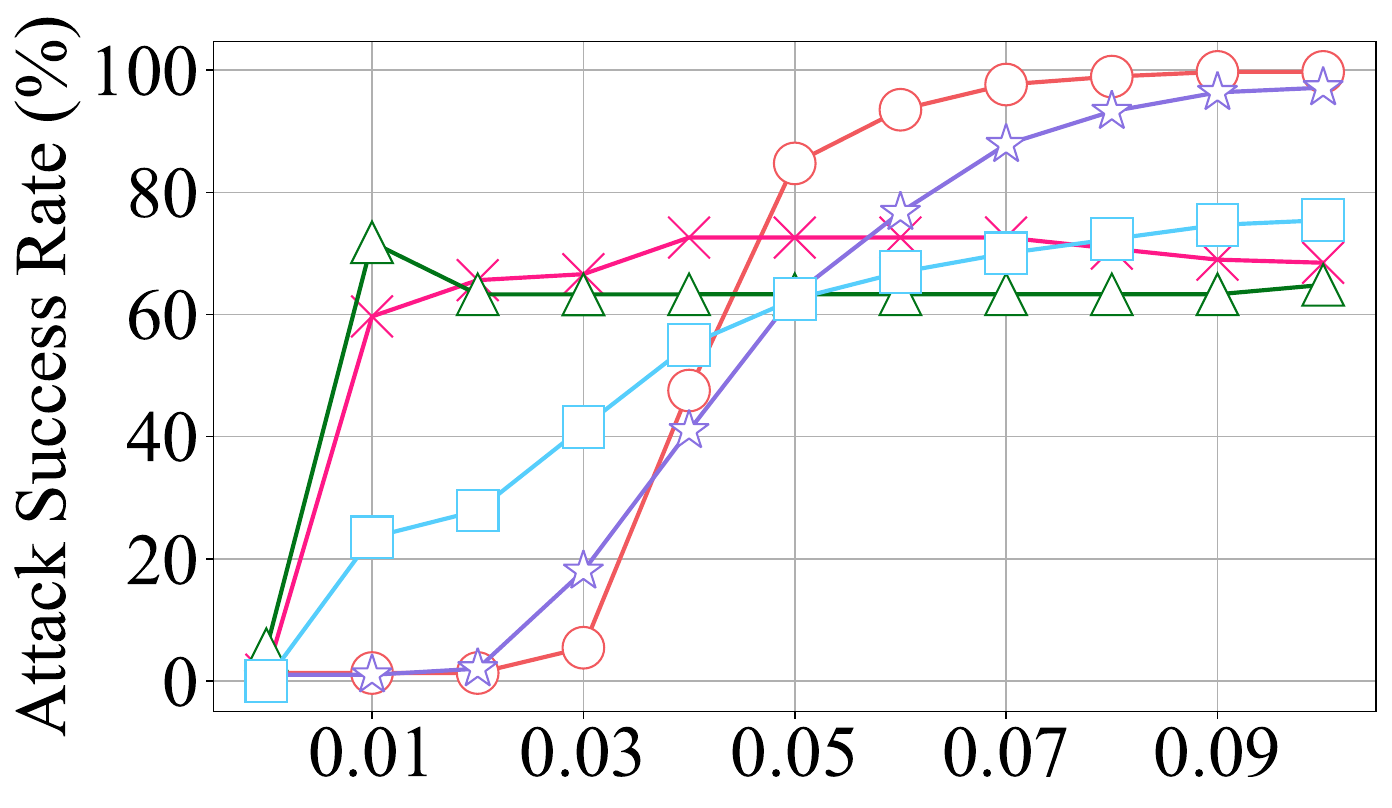}
    \caption{TwitterSpam}
   \end{subfigure}
   \begin{subfigure}{\textwidth}
    \includegraphics[width=\textwidth]{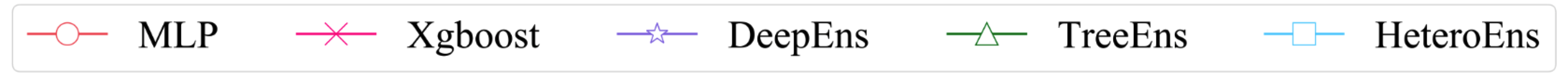}
\end{subfigure}
\caption{The effects of attack cost of different attacks: Deep model changes more smoothly than tree model.}\label{fig-eps}
\end{minipage}
\hfill
\begin{minipage}[b]{0.5\textwidth} 
  \begin{subfigure}{0.5\textwidth}
   \includegraphics[width=\textwidth]{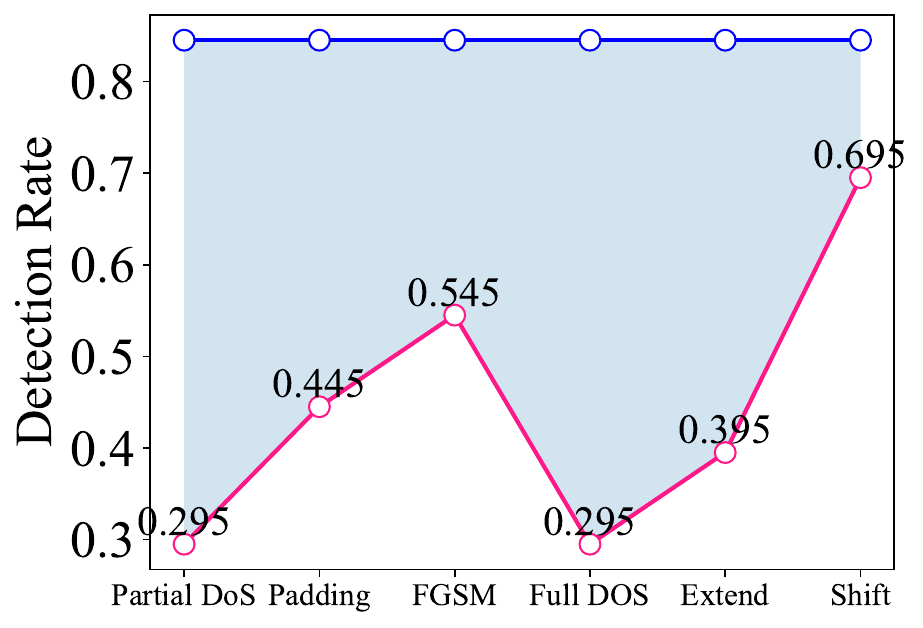}
   \caption{White-box attack}
  \end{subfigure}
  \hfill
  \begin{subfigure}{0.5\textwidth}
   \includegraphics[width=\textwidth]{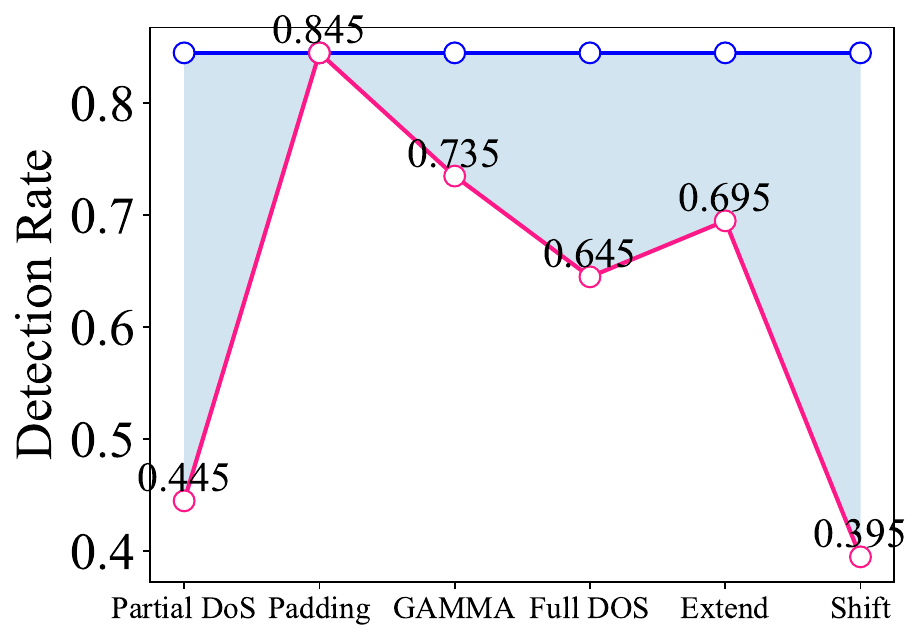}
   \caption{Black-box attack}
 \end{subfigure}
 \caption{The attack effectiveness of End-to-end adversarial attack against MalConv.} \label{fig-end-to-end}
\end{minipage}
\end{figure*}

\subsubsection{Attack Effectiveness Comparison (Different Classifiers)}
In Table \ref{attack-measure}, it can be seen that LR is the most robust gradient model, followed by SVM and then deep learning. 
It is difficult to break the LR model with FGSM, PGD, or C\&W attacks, which achieve only 6.20\% ASR compared to over 90\% success with deep learning (MLP, CNN, RNN, and Transformer).
It is interesting to note that ZOO is highly effective in attacking DT, perhaps because DT uses only one feature as input, while ZOO also utilizes a single feature when evaluating gradients.
In addition, there is another situation that cannot be ignored. 
The tree model (DT, RF, and Xgb.), even though it cannot be attacked by gradient attack, appears vulnerable to these decision-based attack methods (BA, HSJA), completing more than 90\% ASR.

\subsubsection{Attack Effectiveness Comparison (Different Datasets)}

We fully evaluate the CICIDS2017 dataset in Table \ref{attack-measure}. 
But we examine only two ensemble models (DeepEns and HeteroEns) on the other three datasets due to space constraints.
In Fig. \ref{fig-datasets}, we choose three gradient-based attacks and three gradient-free attacks for the DeepEns model, while we can only evaluate six gradient-free attacks for the HeteroEns model.
The four datasets in Fig. \ref{fig-datasets} clearly show that gradient-based attacks outperform gradient-free attacks, and of all gradient-free attacks, HSJA is the clear winner, which is consistent with the conclusion of the CICIDS2017 dataset.

\subsubsection{Attack Effectiveness Comparison (Different Attack Cost)}

The effectiveness of adversarial attacks is closely related to the cost of different attacks, and in this section, we investigate the attack effect of varying attack costs for different models.
Fig.~\ref{fig-eps} shows that for the CICIDS2017 dataset, the deep models such as MLP and DeepEns transform more smoothly for different attack costs, while other models have a sudden change when the attack cost is 0.005.
Displays the same phenomenon on the TwitterSpam dataset.
In contrast, a side-by-side comparison of these two datasets reveals that CICIDS2017 has a lower attack cost than TwitterSpam.

\subsubsection{End-to-end Attack Effectiveness Comparison}

To verify that our framework can scale to end-to-end problem-space attacks, we examine the effectiveness of adversarial attacks in malware detection.
We analyze six white-box attacks and six black-box attacks in Fig.~\ref{fig-end-to-end}. 
The ordinate represents the detection rate. 
In the beginning, the detection rate of the MalConv model was 0.845.
The detection rate of the model decreased after the adversarial attack was launched.  
The strongest Full DOS attack yields the lowest detection rate only achieving 0.295. Side-by-side comparisons of white-box attacks and black-box attacks show that in general, white-box attacks will produce greater damage.

\textbf{Remark 1.} \textit{The gradient-based attacks, under the same conditions, have the stronger attack effect, while HSJA has the best attack effect among all gradient-free attacks.
LR is the most robust model, while the deep model is the least robust.
In fact, even the tree model cannot be defeated to gradient-based attacks, but it is not robust enough. 
Gradient-free attacks can easily break tree models, which are even more vulnerable than deep models.
}

\subsection{Basic Evaluation of Defenses}
\label{sec-ex-def}
In this section, we evaluate the defensive effects of commonly used adversarial defense techniques: model ensembles and adversarial training (AT).

\begin{figure*}
  \centering
  \begin{subfigure}{.49\textwidth}
    \caption{CICIDS2017}
   \includegraphics[width=\textwidth]{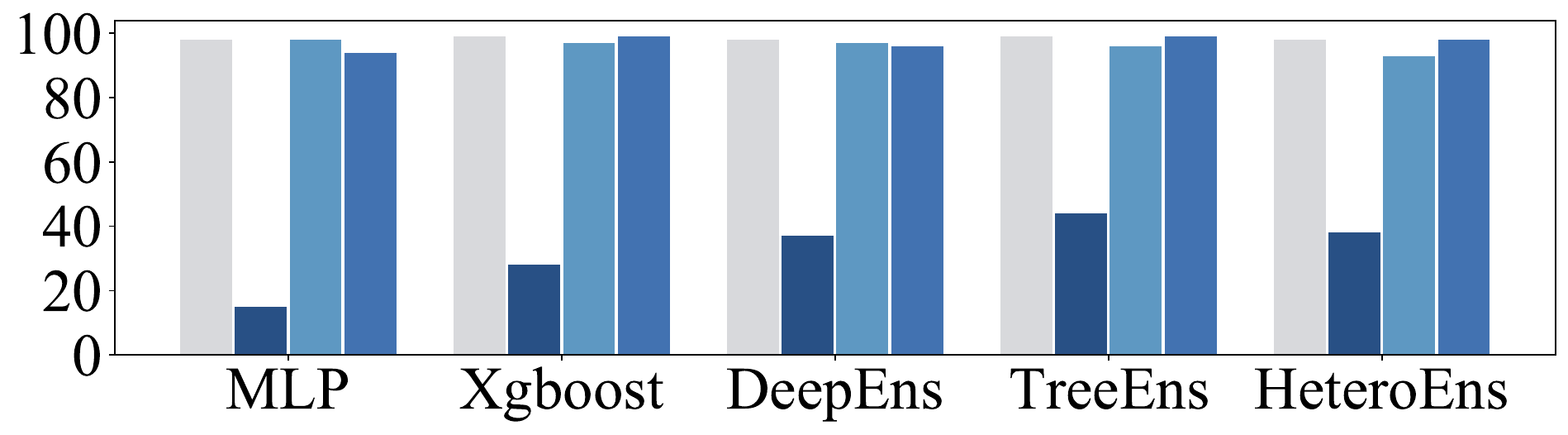}
  \end{subfigure}
  \hfill
  \begin{subfigure}{.49\textwidth}
    \caption{TwitterSpam}
   \includegraphics[width=\textwidth]{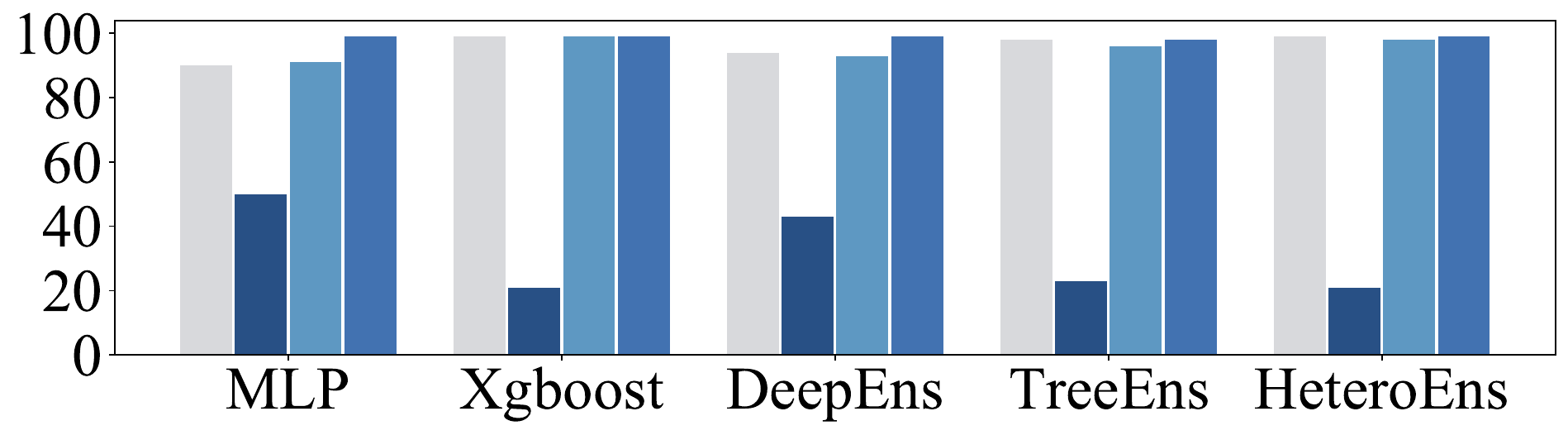}
 \end{subfigure}
 \begin{subfigure}{.49\textwidth}
  \caption{Bodmas}
  \includegraphics[width=\textwidth]{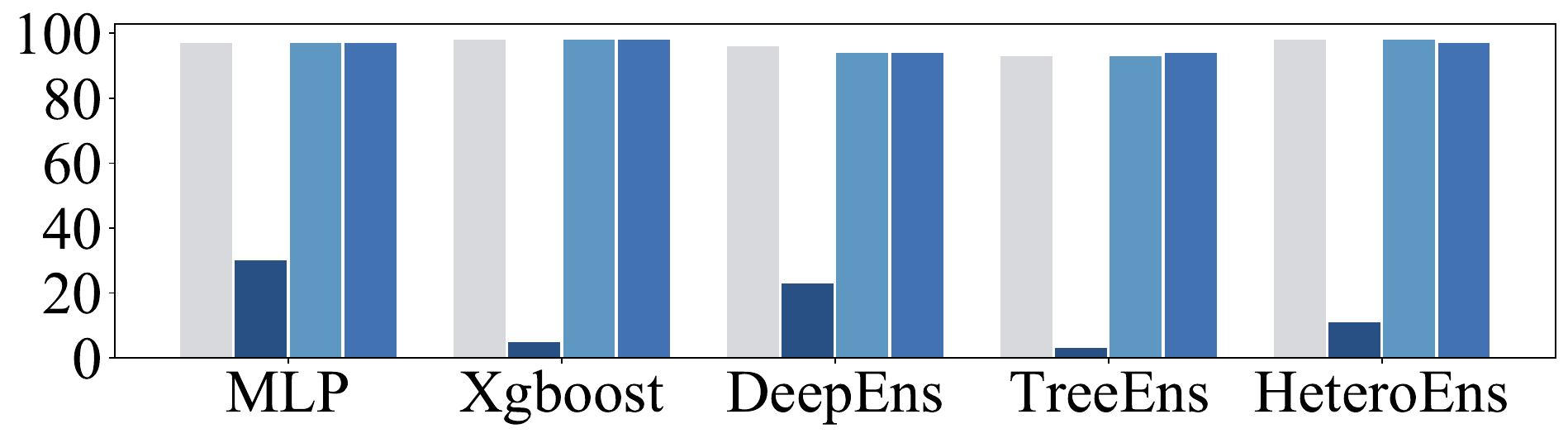}
\end{subfigure}
\begin{subfigure}{.49\textwidth}
  \caption{CICAndMal2017}
  \includegraphics[width=\textwidth]{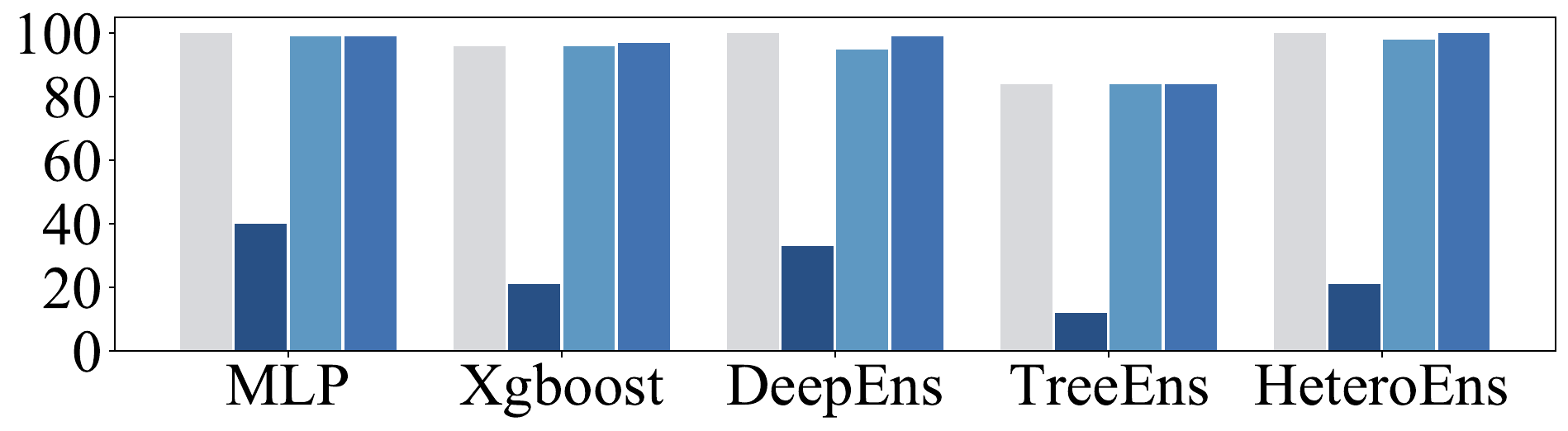}
\end{subfigure}
\begin{subfigure}{0.65\textwidth}
  \includegraphics[width=\textwidth]{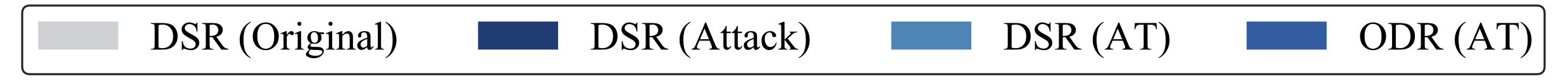}
\end{subfigure}
 \caption{To measure the effectiveness of adversarial training (AT) on four security datasets, we choose five models (MLP, Xgb., DeepEns, TreeEns, and HeteroEns).
          AT increases the defense success rate (DSR) while barely affecting the original detection rate (ODR), demonstrating that it is a mature defense technique.}
 \label{fig-at-effect}
 \vspace{-0.06in}
\end{figure*}

\subsubsection{The Effectiveness of Ensemble Model for Defense}

Earlier studies have indicated that model ensembles can enhance the robustness of the original models, 
and we can see from Table \ref{attack-measure} that model ensemble has a slight advantage over single-model robustness in some cases, 
e.g., DT is completely breached when dealing with ZOO attacks, but ZOO can only achieve 50.85\% ASR with the TreeEns model.
Despite that, these results do not support the conclusion that model ensembles enhance system robustness. 
In fact, under both BA and HSJA attacks, all ensembles except the deep ensemble model achieve an ASR of 96\% or higher.
As shown in Fig. \ref{fig-datasets}, neither DeepEns nor HeteroEns are able to withstand all adversarial attacks on four security datasets.
In Section \ref{sec-adaptive-attacks}, we can clearly see that model ensemble is even more ineffective against stronger attacks-adaptive attacks.

\subsubsection{Adversarial Training Effectiveness}

Adversarial training (AT) is another, more effective method of defense. 
To measure the effectiveness of naive adversarial training (NAT) on four security datasets, we choose five models (MLP, Xgb., DeepEns, TreeEns, and HeteroEns). 
In Fig.~\ref{fig-at-effect}, it is clearly visible that the original detection rate (ODR) drops significantly after either model is attacked, but after AT, the detection rate for the adversarial examples increases significantly right away, reaching over 95\% DSR.
It is also important to note that AT hardly affects the original detection rate (ODR), proving it to be a proven defense technique.
Thus, it begs the question: \textbf{Can the robustness of the model be guaranteed by simply using general adversarial training? 
The answer is no, as detailed in later experiments, where we will describe the limitations of native adversarial training (NAT) and our improvement methods.}

\textbf{Remark 2.} \textit{
The model ensemble cannot theoretically guarantee robustness, and our experiments demonstrate that neither the DeepEns nor the HeteroEns can withstand all adversarial attacks.
As a more effective defense method, adversarial training (AT) increases the defense success rate (DSR) while barely affecting the original detection rate (ODR), demonstrating that it is a mature defense technique.
NAT still has limitations: the robust generalization ability is weak, so it can only resist a single attack method.
}
\subsection{Evaluation of Adaptive Attacks}
\label{sec-adaptive-attacks}
In this section, we evaluate adaptive attacks in two scenarios: 1) \textit{ensembles of multiple attacks (EMA)}, and 2) \textit{transfer ensemble attacks (TEA)}.
\begin{figure*}
  \begin{minipage}[b]{0.5\textwidth} 

  \begin{subfigure}{\textwidth}
    \caption{CICIDS2017}
   \includegraphics[width=\textwidth]{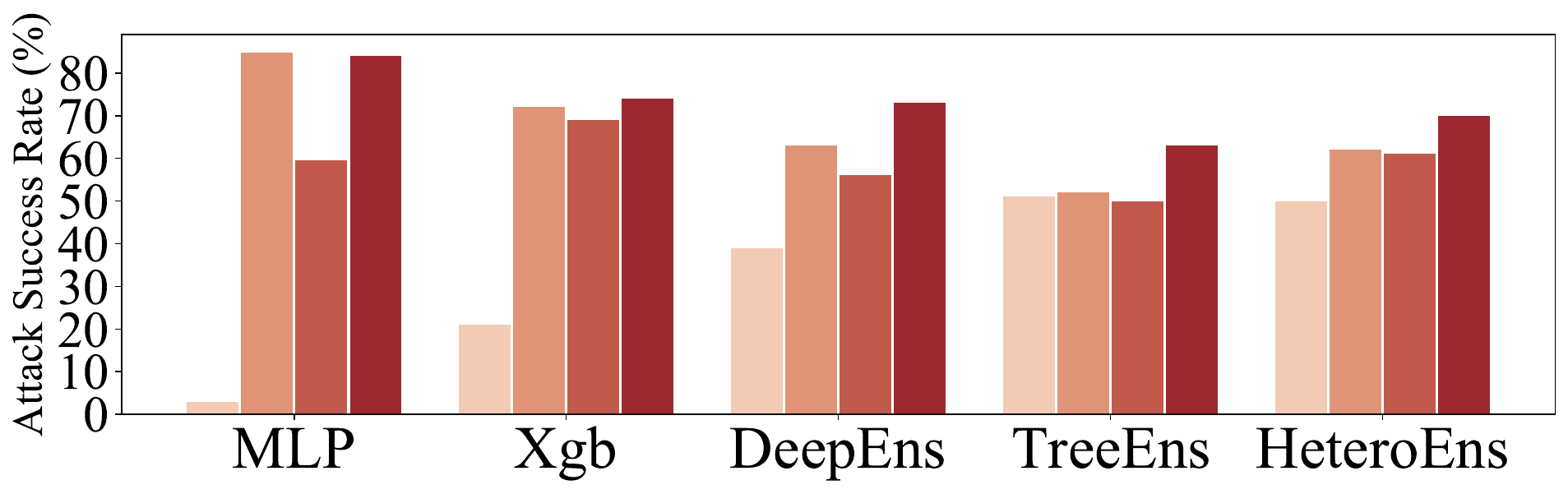}
  \end{subfigure}

  \begin{subfigure}{\textwidth}
    \caption{TwitterSpam}
   \includegraphics[width=\textwidth]{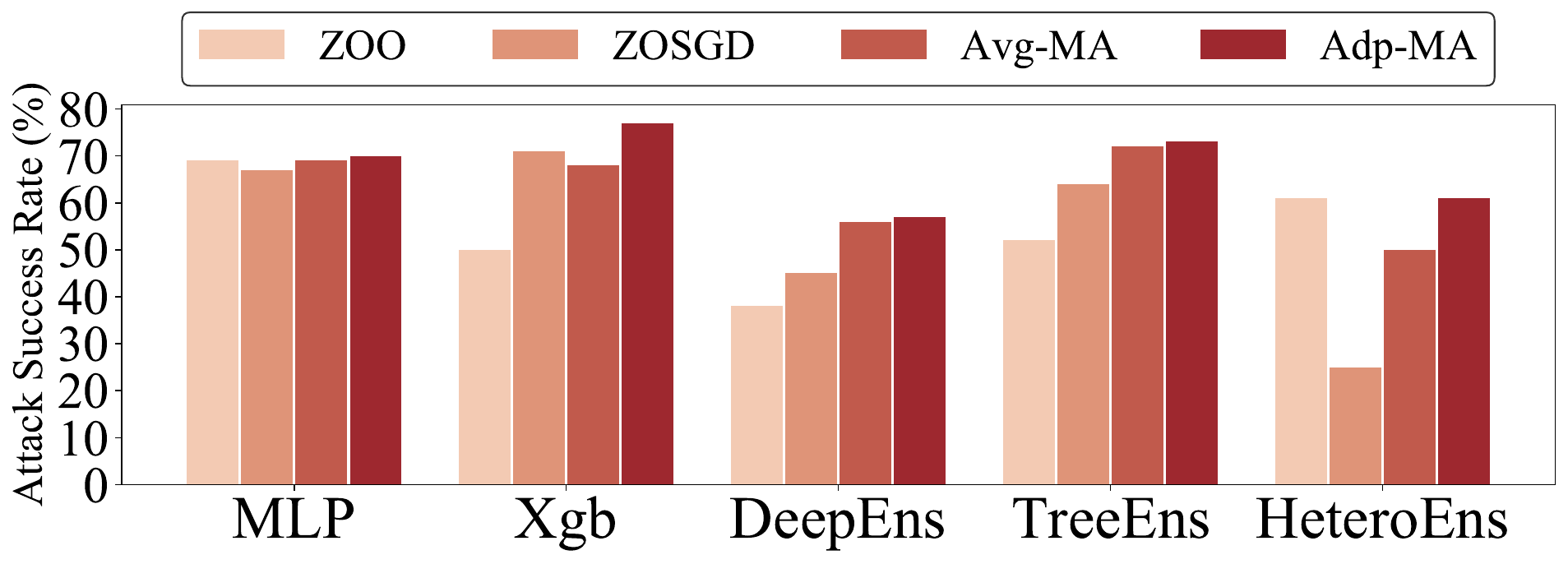}
 \end{subfigure}
 \caption{A comparison of different \textit{Ensemble of Multiple Attacks} methods: Attack performance is not improved by Avg-EA, but Adp-EA can significantly improve attack performance.} \label{adp-ens-attack}
 \end{minipage}
 \hfill
 \begin{minipage}[b]{0.5\textwidth} 

 \begin{subfigure}{\textwidth}
  \caption{CICIDS2017}
  \includegraphics[width=\textwidth]{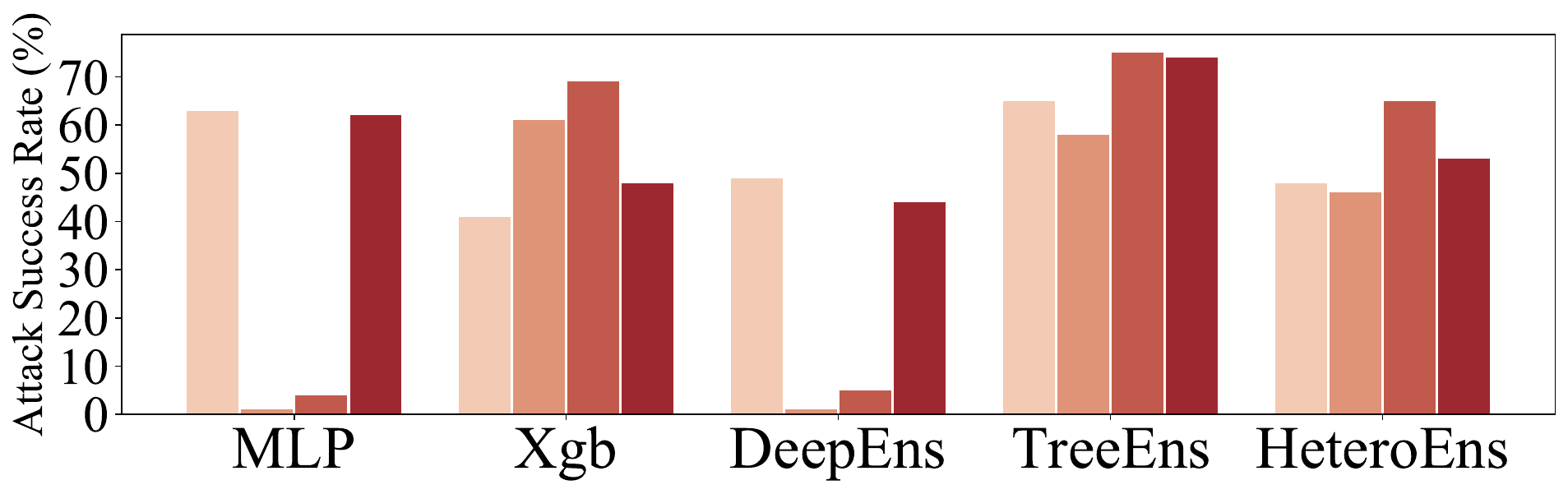}
\end{subfigure}

\begin{subfigure}{\textwidth}
  \caption{TwitterSpam}
  \includegraphics[width=\textwidth]{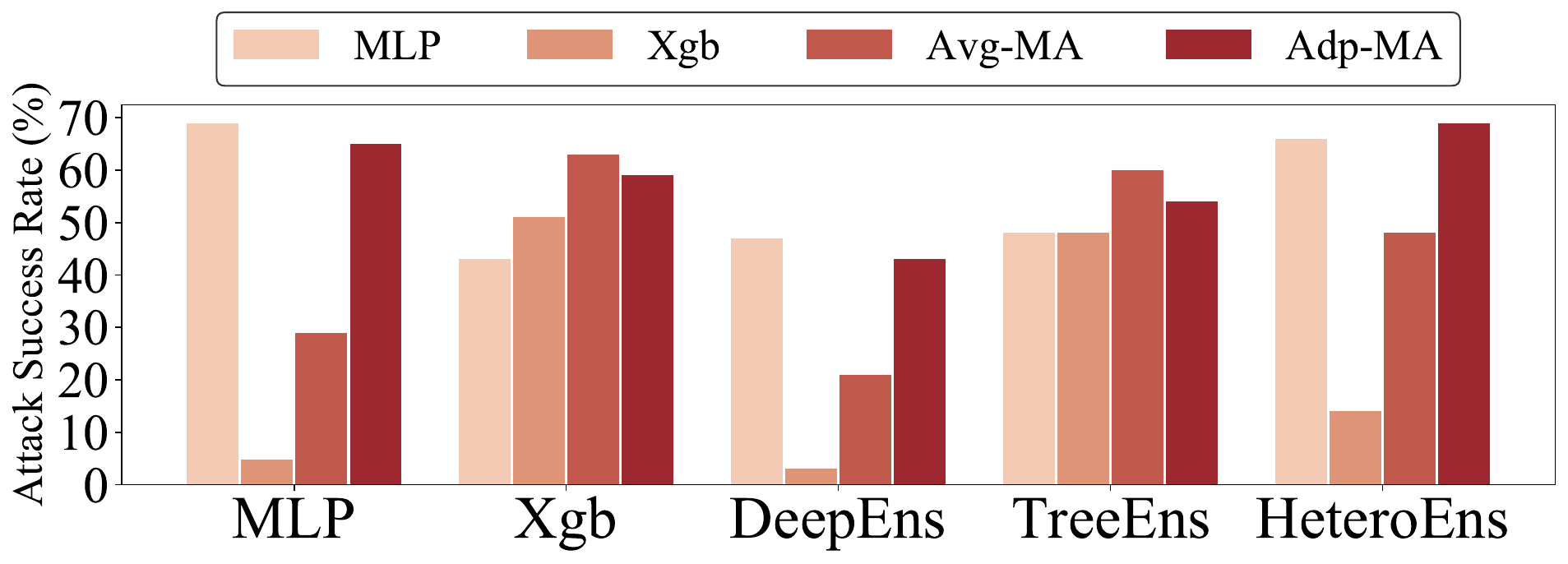}
\end{subfigure}
\caption{A comparison of different \textit{transfer ensmeble attacks}: An adaptive transfer ensemble attack (Adp-EA) yields better results than an average transfer ensemble attack (Avg-EA). } \label{adp-trans-attack}
\end{minipage}
\end{figure*}
\begin{figure*}
  \centering
  \begin{subfigure}{.49\textwidth}
    \caption{CICIDS2017}
    \includegraphics[width=\textwidth]{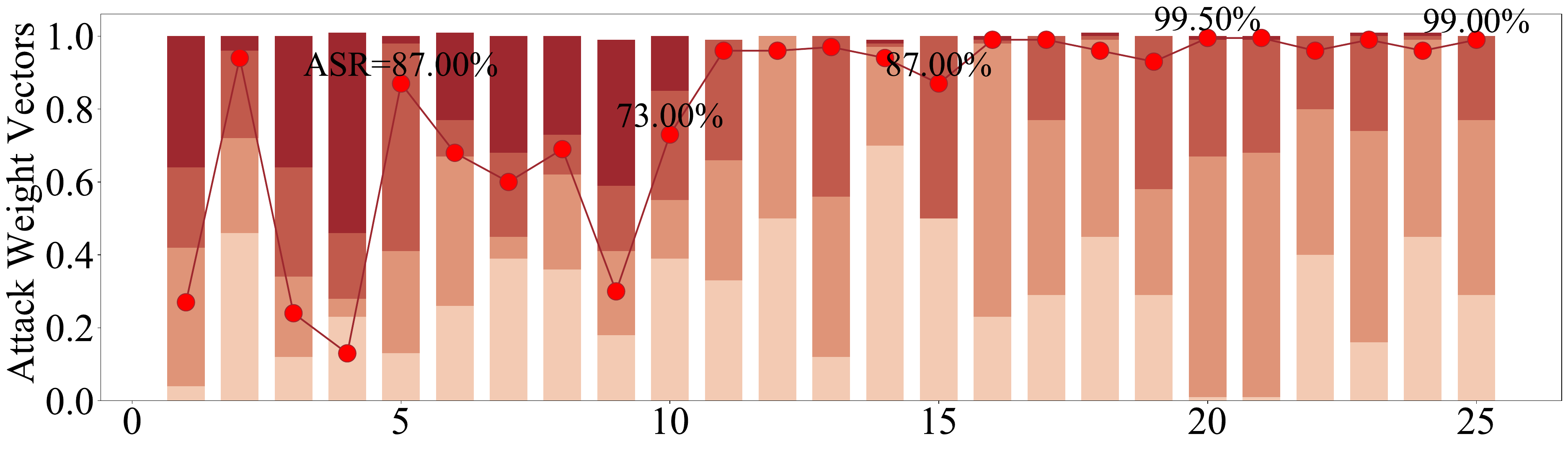}
  \end{subfigure}
  \hfill
  \begin{subfigure}{.49\textwidth}
    \caption{TwitterSpam}
    \includegraphics[width=\textwidth]{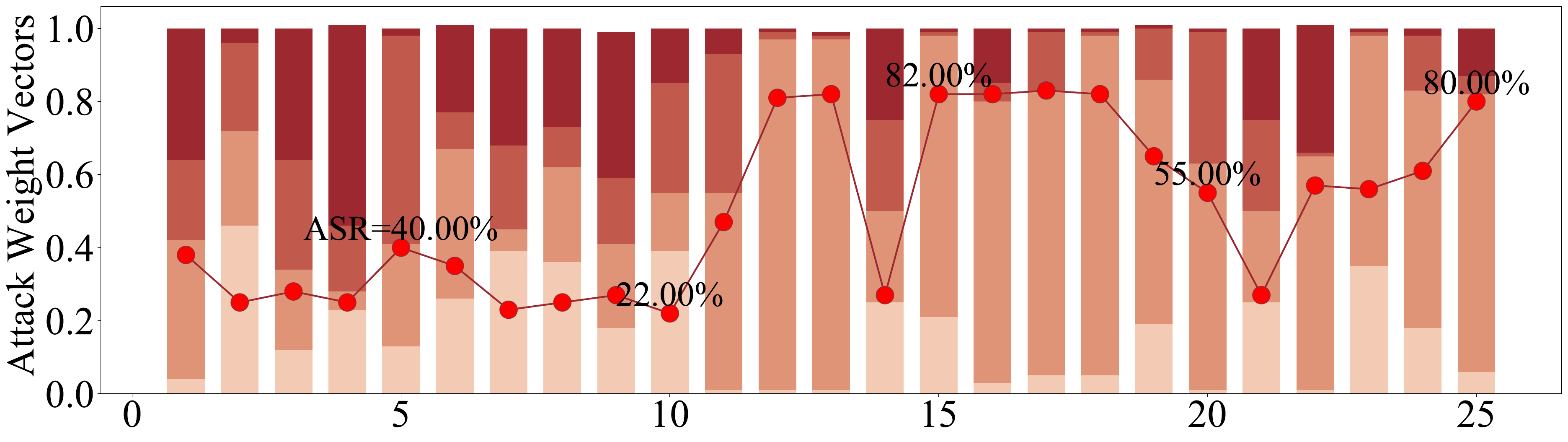}
    \end{subfigure}
  \begin{subfigure}{0.45\textwidth}
    \includegraphics[width=\textwidth]{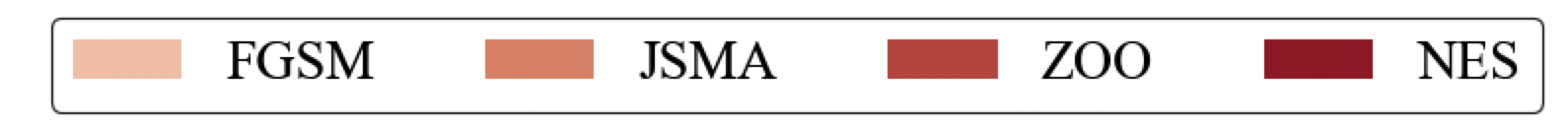}
  \end{subfigure}
  \caption{The comparison of the attack effect of different ensemble adversarial attack weights for ensemble attack methods.} \label{fig-attack-eps}
  \vspace{-0.06in}
\end{figure*}

\subsubsection{Ensemble of Multiple Attacks}
\textit{Ensemble of Multiple Attacks (EMA)} represents the ensemble of multiple attack methods (e.g. FGSM, JSMA and ZOO).
As shown in Fig.~\ref{adp-ens-attack}, we evaluate the robustness of five models when faced with adaptive attacks, including DeepEns, TreeEns and HeteroEns models.
In view of the fact that three of the models (Xgb, TreeEns, and HeteroEns) are not differentiable, we choose two gradient-free attacks to assess the adaptive attacks' efficiency.
On the CICIDS2017 dataset, the ZOO attack is less effective than ZOSGD on all models, but on the TwitterSpam dataset, ZOO is more effective than ZOSGD on both MLP and HeteroEns models, especially the HeteroEns model, which is about 40\% better than ZOSGD.

And then, we measure the attack effectiveness of the Average Ensemble Attacks (Avg-EA).
The Avg-EA is a compromise between ZOO and ZOSGD, and it is about 10\% worse than ZOSGD on the CICIDS2017 dataset.
In contrast, on the TwitterSpam dataset, Avg-EA outperforms the single attack approach by 10\% to 15\% on both DeepEns and TreeEns models.
There is an interesting observation that when both attack methods of Avg-EA are equally effective, Avg-EA is better, with Avg-EA on the TwitterSpam dataset being more stable and stronger than on CICIDS2017.
As for our adaptive ensemble adversarial attack (Adp-EA), it outperforms the other attack methods significantly on all datasets, and outperforms the Avg-EA by 15\%, especially when the Avg-EA attack method performs poorly. 
On the other hand, all ensemble models cannot guarantee model robustness against adaptive adversarial attacks, and adaptive attacks on all datasets can achieve more than 50\%  ASR against ensemble models.

\subsubsection{Transfer Ensemble Attacks.}
Next, we evaluate the effectiveness of the transfer ensemble attack (TEA).
We attack the five classifiers using MLP, Xgb, Avg-MA and Adp-MA as substitute models, respectively.
As shown in Fig.~\ref{adp-trans-attack}, attacking MLP with MLP as a substitute model is the most effective in the CICIDS2017 dataset, but Avg-EA has the best attack effect against Xgb.
On the other hand, Avg-EA performs well when attacking Xgb, TreeEns, and HeteroEns models, and its $ASR$ is 60\%, but it performs poorly against MLP and DeepEns models, and its $ASR$ is less than 10\%.
It is for this reason that we study adaptive attacks. 
While not as effective on some models as Avg-EA, our Adp-EA performs consistently, with an $ASR_{avg}$ of over 50\%.
Performance on the TwitterSpam dataset is similar to that on CICIDS2017.

\subsubsection{The comparison of different attack weights for ensemble attack methods.}
Finally, we evaluate the adaptive ensemble attack (Adp-EA) using Bayesian optimization for weights change using four attack methods (FGSM, JSMA, ZOO and NES) on two datasets. 
Bayesian optimization has a very high optimization efficiency. 
In 15th searches, Bayesian optimization has almost reached a steady state in the CICIDS2017 dataset.
According to Fig.~\ref{fig-attack-eps}, the optimal result is obtained on the 19th attempt, whose ratio of JSMA and ZOO was 7:3, and the attack success rate (ASR) reached 99.5\%.
For the TwitterSpam dataset, the optimal result occurs at the 15th, when the proportion of FGSM and JSMA is relatively high.

\textbf{Remark 3.} \textit{In either case (EMA or TEA), an adaptive ensemble attack (Adp-EA) is more effective than a single attack or ensemble attack. 
In addition, our Bayesian optimization-based Adp-EA is highly efficient, making it suitable for efficient evaluation. 
Moreover, the ensemble model (TreeEns, DeepEns and HeteroEns) is more hopeless when faced with adaptive attacks.
}

\subsection{Multi-Attack Adversarial Training.}
\label{sec-ex-ma-at}

\subsubsection{Multi-Attack Adversarial Training.}
\begin{figure*}
  \centering
  \begin{subfigure}{.26\textwidth}
    \caption{CICIDS2017}
   \includegraphics[width=\textwidth]{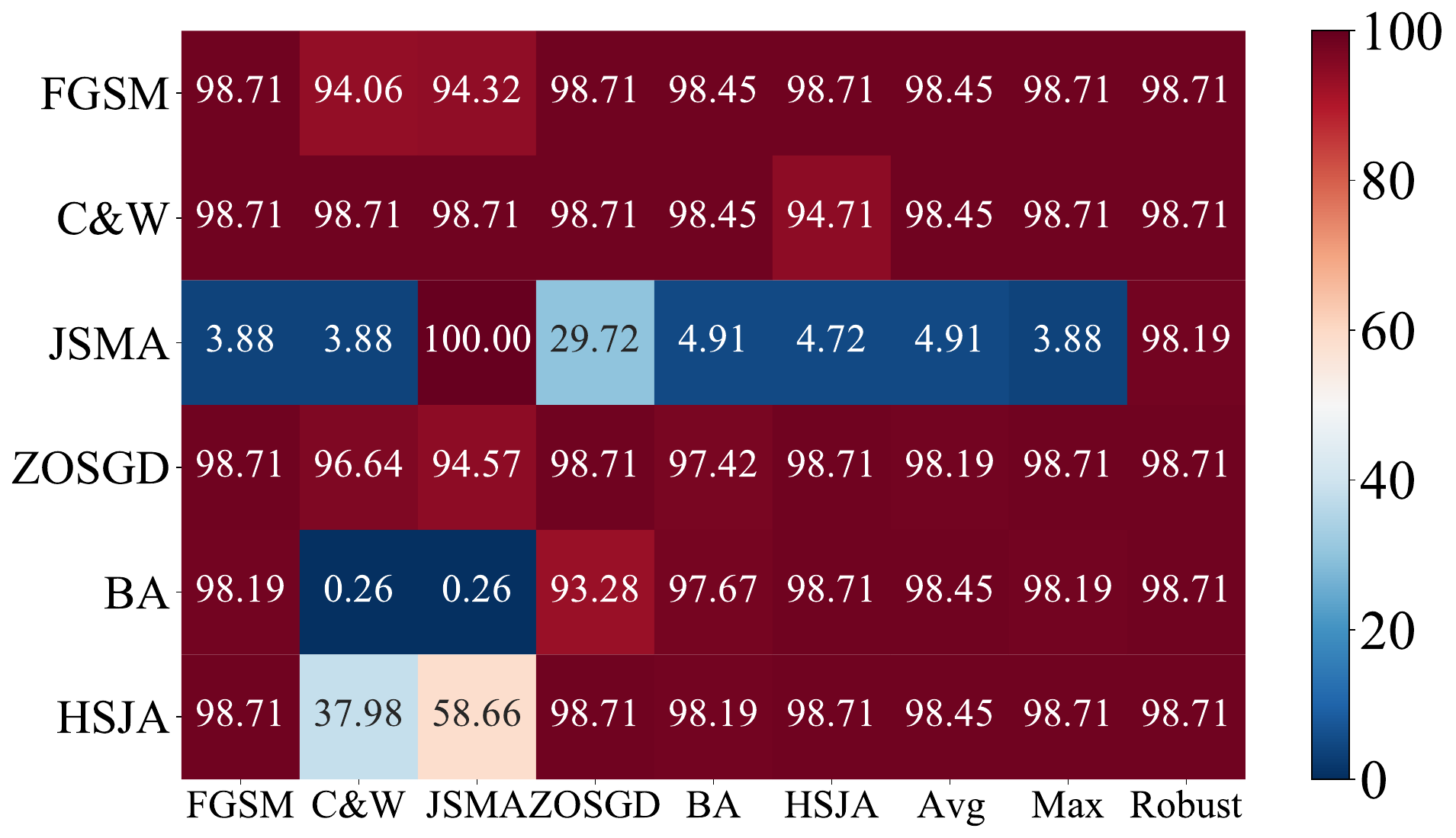}
  \end{subfigure}
  \hfill
  \begin{subfigure}{.24\textwidth}
    \caption{Twitter Spam}
   \includegraphics[width=\textwidth]{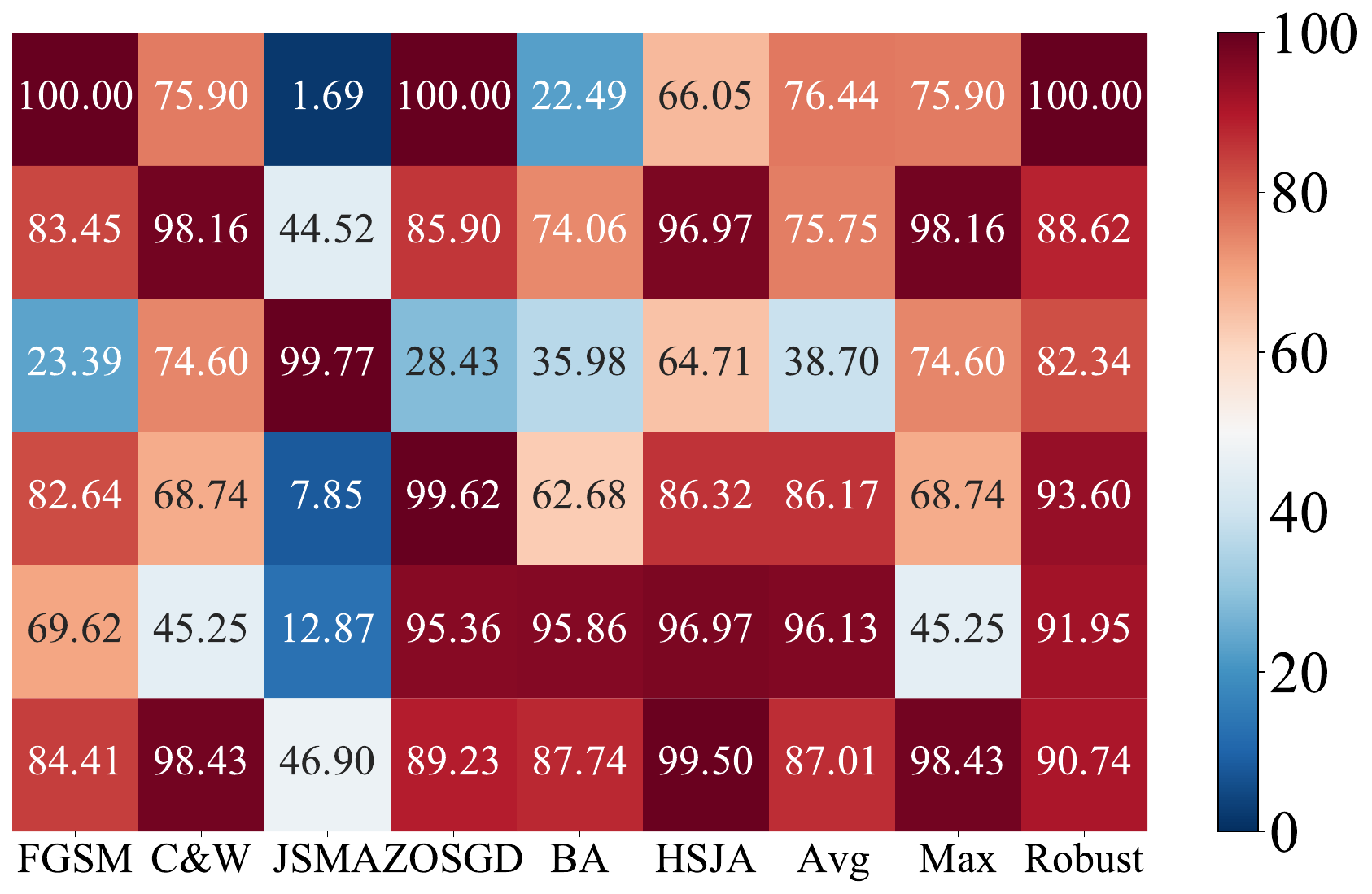}
 \end{subfigure}
 \hfill
  \begin{subfigure}{.24\textwidth}
    \caption{Bodmas}
    \includegraphics[width=\textwidth]{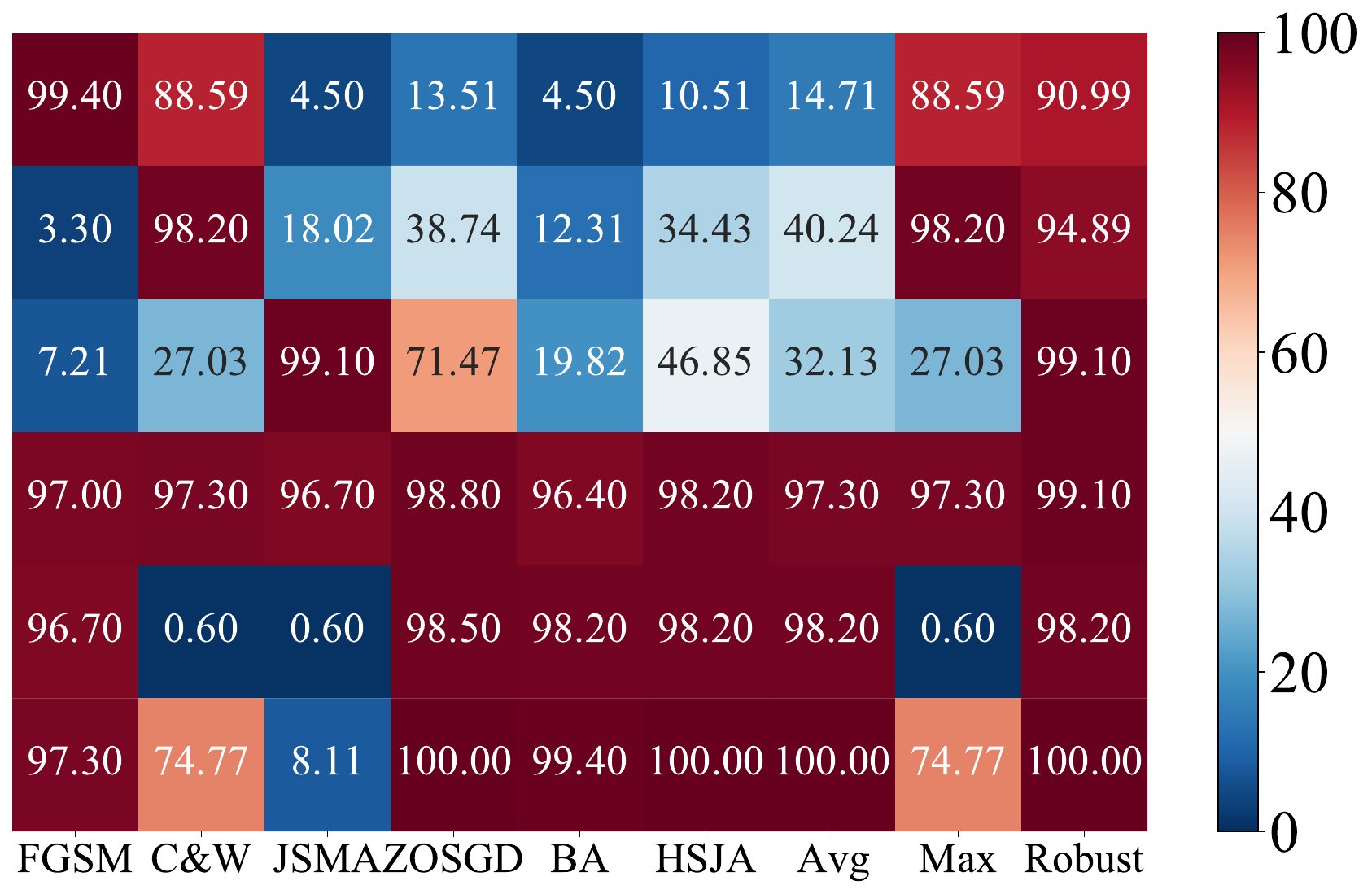}
   \end{subfigure}
   \hfill
 \begin{subfigure}{.24\textwidth}
  \caption{CICAndMal2017}
 \includegraphics[width=\textwidth]{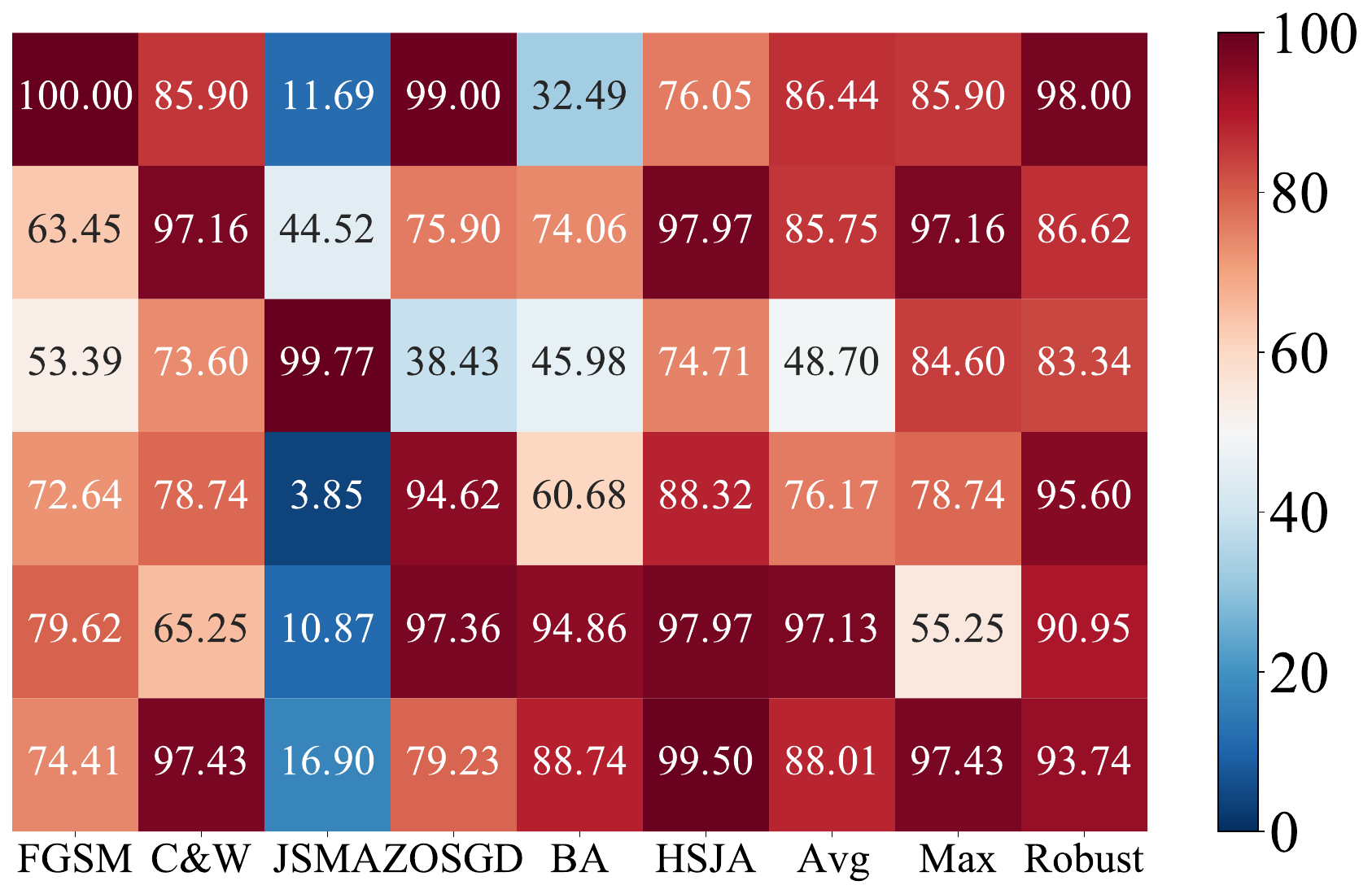}
\end{subfigure}
 \caption{Cross-attack transferability matrix: Cell $(i, j)$ represents the robust model produced by using attack $i$ to attack the AT method $j$. The rows indicate the attack methods that craft AEs, and the columns indicate the models under AT.}
 \label{fig-AT-trans}
 \vspace{-0.06in}
\end{figure*}
\begin{table*}[]
  \caption{The Multi-attack adversarial training effectiveness (DSR \%)}
  \label{table-multi-AT} 
  \begin{center}
  \begin{tabular}{@{}ccccccccccccc@{}}
  \toprule
  Dataset & \multicolumn{3}{c}{CICIDS2017} & \multicolumn{3}{c}{TwitterSpam} & \multicolumn{3}{c}{Bodmas} & \multicolumn{3}{c}{CICAndMal2017} \\ \midrule
  \diagbox{Attack}{Defense}  & Avg-AT (\%)   & Max-AT   & R-AT   & Avg    & Max   & Robust   & Avg  & Max  & Robust & Avg    & Max    & Robust    \\ \hline
  FGSM    & 98.45    & 98.71    & 98.71    & 76.44     & 75.9     & 100      & 14.71   & 88.59   & 90.99  & 86.44     & 85.9      & 98        \\
  C\&W    & 98.45    & 98.71    & 98.71    & 75.75     & 98.16    & 88.62    & 40.24   & 98.2    & 94.89  & 85.75     & 97.16     & 86.62     \\
  JSMA    & 4.91     & 3.88     & 98.19    & 38.7      & 74.6     & 82.34    & 32.13   & 27.03   & 99.1   & 48.7      & 84.6      & 83.34     \\
  ZOSGD   & 98.19    & 98.71    & 98.71    & 86.17     & 68.74    & 93.6     & 97.3    & 97.3    & 99.1   & 76.17     & 78.74     & 95.6      \\
  BA      & 98.45    & 98.19    & 98.71    & 96.13     & 45.25    & 91.95    & 98.2    & 0.6     & 98.2   & 97.13     & 55.25     & 90.95     \\
  HSJA    & 98.45    & 98.71    & 98.71    & 87.01     & 98.43    & 90.74    & 100     & 74.77   & 100    & 88.01     & 97.43     & 93.74     \\ \hline
  Adp-EA  & 4.91     & 3.88     & \textbf{98.19}    & 38.7      & 45.25    & \textbf{82.34}    & 14.71   & 0.6     & \textbf{90.99}  & 48.7      & 55.25     & \textbf{83.34}     \\ \bottomrule
  \end{tabular}
\end{center}
  \end{table*}

  \begin{figure*}
    \centering
    \begin{subfigure}{.49\textwidth}
      \caption{CICIDS2017}
     \includegraphics[width=\textwidth]{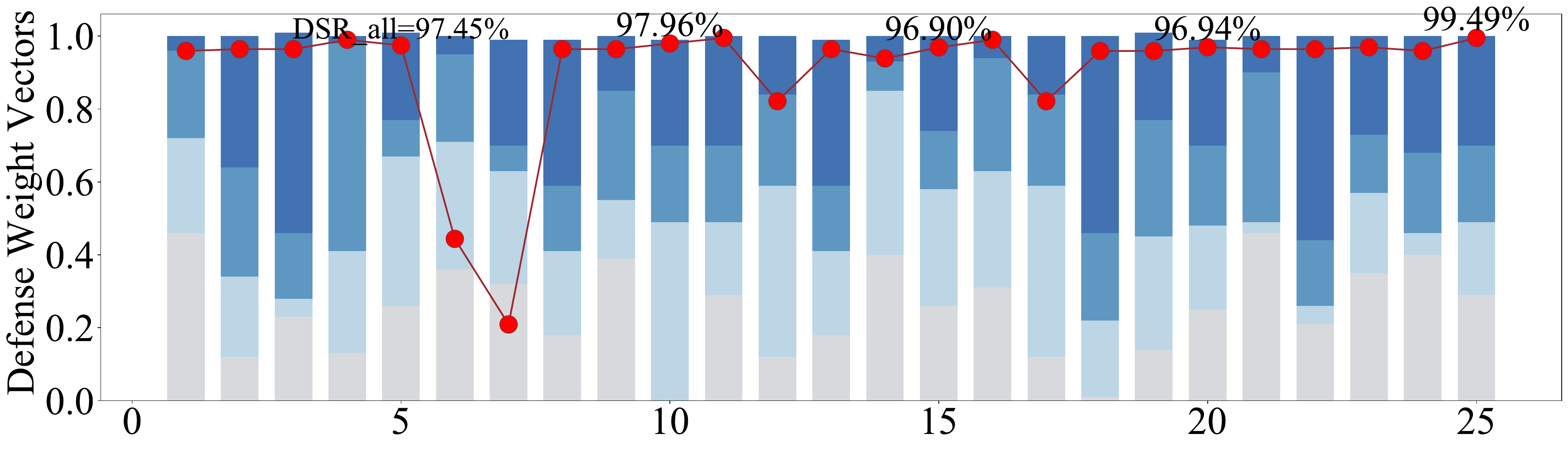}
    \end{subfigure}
    \hfill
    \begin{subfigure}{.49\textwidth}
      \caption{TwitterSpam}
      \includegraphics[width=\textwidth]{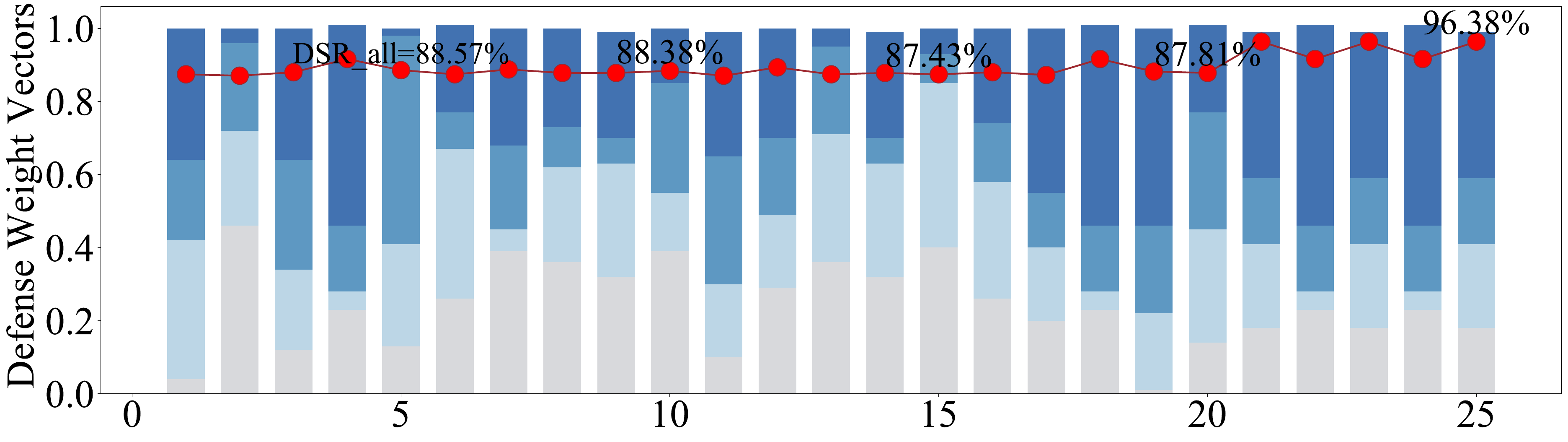}
     \end{subfigure}
    \begin{subfigure}{0.45\textwidth}
      \includegraphics[width=\textwidth]{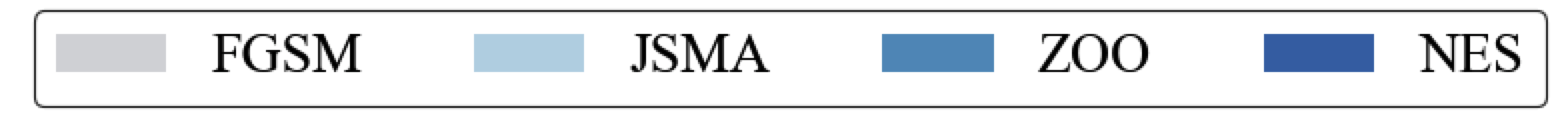}
    \end{subfigure}
   \caption{The comparison of the defense effect of different ensemble AT weights for EMA.} \label{fig-at-w}
   \vspace{-0.06in}
  \end{figure*}

We explore the \textit{robust generalization} abilities of the model resulting from adversarial training (AT) involving a variety of adversarial attacks in Fig.~\ref{fig-AT-trans}.
We choose three white-box attacks (FGSM, C\&W and JSMA) and three black-box attacks (ZOSGD, BA and HSJA) as the attacks to generate adversarial examples (AEs).
In addition to selecting these six single attacks as methods of generating AEs for AT, we also select three ensemble adversarial training strategies.
We built a cross-attack transferability matrix, where each cell $(i, j)$ represents the robust model produced by using attack $i$ to attack the AT method $j$.
Specifically, the rows represent the attack method used and the columns represent the AT method.
It is most evident in the TwitterSpam dataset that using the same attack method for both the AT and the adversarial attack is most effective, resulting in over 90\% DSR.

The FGSM performs best in all AT using a single attack, as in the CICIDS2017 dataset, reaching a DSR of 98\% in all attacks except for the attack method using JSMA.
TwitterSpam dataset illustrates the robustness of AT that is difficult to transfer against other adversarial attack i.e., less robust generalization.
Three ensemble adversarial training methods: Avg-AT, Max-AT and R-AT have better results compared to other native adversarial training methods.
However, Avg-AT and Max-AT do not achieve good robustness against certain attacks.
For example, in the Bodmas dataset, Avg-AT only achieves about 30\% DSR against FGSM, C\&W and JSMA attacks, and Max-AT only achieves 0.6\% DSR against BA attacks.
In contrast, our robust AT methods (R-AT) are able to achieve stable robustness against all attacks. 
Specifically,  R-AT achieves 98.71\% DSR in dataset CICIDS2017, more than 90\% DSR in Bodmas, and more than 80\% DSR in TwitterSpam, showing that our robust training approach can meet the robustness in real situations.

In our final evaluation, we examine whether general multi-attack adversarial training can meet the robustness requirements when faced with adaptive attackers.
As shown in Table~\ref{table-multi-AT}, we found that when faced with Adp-EA in CICIDS2017, the average multi-attack adversarial training (Avg-AT) and maximum adversarial training (Max-AT) both reached 4.91\% and 3.88\% DSR, respectively, which means that the two defense effects are completely invalid, while our robust adversarial training (R-AT) achieves 98.19\% DSR.
Other datasets show similar patterns as CICIDS2017, but the difference between Robust strategies on TwitterSpam and CICAndMal2017 is less obvious than on other datasets.
In conclusion, our robust ensemble adversarial training method remains a successful defense strategy.

\subsubsection{The comparison of different defense weights for multi-Attack adversarial training.}
We choose an adaptive ensemble of four basic attacks (FGSM, JSMA, ZOO and NES) methods, and then assess the effectiveness of three adversarial trainings that are based on multiple attacks.
In Fig.~\ref{fig-at-w}, the proportion of weights is relatively stable.
On the CICIDS2017 dataset, the proportions of the four base kinds of adversarial training are relatively uniform. 
When the proportion of FGSM and NES is slightly higher, the best defense effect obtains, and the DSR equals 99.49\%.
In the TwitterSpam dataset, the changes of different adversarial training methods are relatively gentle, and the more NES is used, the better the defense effect is.

\textbf{Remark 4.} \textit{
In the face of multiple attack methods, the average ensemble adversarial training (Avg-AT) and the maximize ensemble adversarial training Max-AT do not achieve good robustness to some attacks.
In contrast, our robust ensemble adversarial training (R-AT) approach achieves stable robustness against all attacks, as well as adaptive attacks.
}

\subsection{Transfer Ensemble Adversarial Training.}
\label{sec-ex-teat}
\begin{figure}
  \centering
   \includegraphics[width=0.45\textwidth]{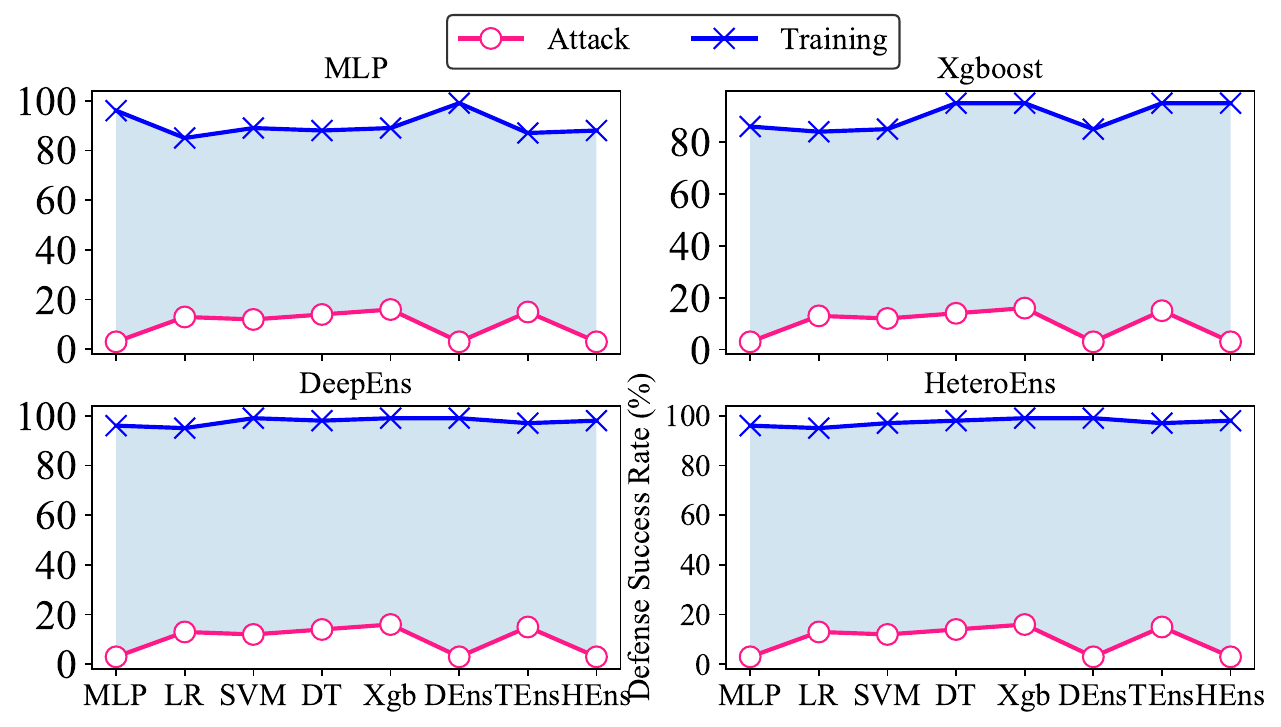}
 \caption{Transfer Ensemble Adversarial Training.} \label{fig-trans-ens-at}
 \vspace{-0.06in}
\end{figure}
Next, we explore the adversarial training methods against the ensemble transfer attack.
We employ adaptive ensemble transfer attack as the attack method to attack eight models (e.g., MLP, Xgb, DeepEns and HeteroEns) and then apply four models (MLP, Xgb, DeepEns and HeteroEns) as the substitute model of adversarial training.
As shown in Fig.~\ref{fig-trans-ens-at}, using different adversarial training substitute models, they all reach high robustness, about 90\% in $DSR_{avg}$.
As a result, the adversarial training method using an ensemble substitute is more effective than using a single model as the substitute model. 
For example, the transfer adversarial training method using the MLP as the substitute achieves 80\% DSR, whereas the transfer adversarial training method using DeepEns as a substitute model achieves more than 95\% DSR.
Therefore, the \textit{transfer ensemble adversarial training (TE-AT)} is already able to achieve high robustness against transfer attacks of different models without using robust adversarial training methods.

\textbf{Remark 5.} \textit{
  Transfer ensemble adversarial training (TE-AT) is more effective than adversarial training with a single model as a surrogate model, and TE-AT already has high robustness to transfer-based attacks on different models.
  Therefore, there is no need to use robust ensemble adversarial training methods.
}

\section{discussion}
We discuss the limitations and some interesting future work in this section.

\noindent\textbf{Limitations.}
First, in order to more quickly validate the robustness of model ensembles and ensemble defenses, we employ a more common and lightweight feature-space instead of a complex end-to-end problem-space attack. 
But there are some security detectors with more sophisticated feature extractors (e.g., packet-based traffic feature extractors-AfterImage \cite{mirsky2018kitsune}),  it is difficult to fully replace problem-space attacks with restricted feature-space attacks.
Further, we improve the robust generalization ability by using unique optimization objectives based on Bayesian optimization, but only show that our method outperforms simple averaging and maximization strategies, which are not theoretically reliable.
Additionally, our logic of "small perturbations" cannot be applied to many security features, which have a variety of costs to manipulate. 
The costs between features are asymmetric and cannot be accurately captured by our $L_p$-norm-based cost model.
The above limitations provide further opportunities for our future work below.

\noindent\textbf{Future Work.}
We also use End-to-end adversarial attacks against PE malware detection in our work to demonstrate how easily our framework can be extended to real-world attacks, and how our gradient-free optimization system is very suitable for discrete security data. 
Then, we will use a complete problem-space attack to demonstrate how superior our framework is.
Additionally, it is common for learners and adversaries to engage in an arms race. 
In a competitive setting, one player's actions are often met with more sophisticated and advanced responses from the other. 
Thus, we need to develop robust generalization capabilities based on game theory with a more theoretical guarantee.
To overcome this last limitation, we will develop a new approach that converts domain knowledge about features into a cost-driven constraint.
In summary, we hope our framework will provide benchmarking platforms for ensemble adversarial attacks and defenses of security detectors, similar to penetration testing platforms applied in the field of cybersecurity.
\section{Conclusion}
This paper presents a first general Cybersecurity Adversarial Robustness Evaluation framework (CARE), which evaluates the impact of model ensembles and ensemble defenses across the entire machine learning space.
The framework is scalable, implements a unified interface for deep learning and machine learning, supports state-of-the-art attacks and defense algorithms, and represents constraints of the problem space in the security domain by remapping functions.
To meet the measurement needs of adaptive attacks in security scenarios, we present two adaptive ensemble attack methods: an adaptive ensemble of multiple attack methods and an adaptive transfer ensemble attack.
Furthermore, we develop and demonstrate our robust ensemble adversarial training, which can withstand multiple attack methods and adaptive attacks simultaneously.
The objective of this paper is to provide a comprehensive and systematic understanding of the fundamental issues of ensemble adversarial attacks and defenses against security detection, thus serving as a key point for the advancement of this research field.

\appendices

\bibliographystyle{IEEEtran}
\bibliography{references}

\end{document}